\documentclass[11pt]{article}
\usepackage[a4paper,top=2.4cm,bottom=2.4cm,left=2.2cm,right=2.2cm]{geometry}
\usepackage{amsmath,amssymb,bm}
\usepackage{graphicx}
\usepackage{booktabs}
\usepackage{array}
\usepackage{caption}
\usepackage[hidelinks]{hyperref}

\title{\bfseries Link prediction in complex networks via fusing node centrality and local similarity indices: a fair-protocol reassessment and parameter design principles}
\author{Yingying Zhang\textsuperscript{1}\quad Chengye Zhao\textsuperscript{1,*}\\[4pt]
{\small 1. College of Sciences, China Jiliang University, Hangzhou 310018, China}\\
{\small * Corresponding author. E-mail: cyzhao@cjlu.edu.cn}\\
{\small ORCID: Y. Zhang 0009-0001-9073-2525; C. Zhao 0000-0002-1149-4419}}
\date{}

\begin{document}
\maketitle

\begin{abstract}
\noindent We reassess a published fusion method under a fair protocol. Local similarity indices are cheap but score every node pair without common neighbors zero, and weighting them by a node centrality product, as proposed for PageRank, is not robust once the protocol is made comparable across indices; on our eight-network panel the reported improvement does not survive that protocol for six of the seven local indices. We study these issues in a unified piecewise framework that modulates the local score multiplicatively by a centrality product when local information suffices and uses a small-dose centrality product as a completion term when it does not. A scale analysis and a reassessment on eight real-world networks (five-fold cross-validation, exhaustive negative samples) trace this to the $O(n^{-2})$ scale of the PageRank product, whose modulation factor is then a near-identity map, and to its rank correlation with degree exceeding $0.89$; control arms show scale calibration is necessary but not sufficient. The min-max normalized DomiRank product instead has an $O(1)$ dynamic range and substantial non-degree content, and a grid search yields unified parameters ($\omega=5$, $c=0.1$) under which DR-JC/HPI/SO significantly outperform their local baselines already at the dataset level ($p=0.027$ and $0.035$), the remaining four at the fold level only; DR-RA attains an average AUC of $0.9257$, above nine of ten comparison methods, whereas degree-only preferential attachment reaches only $0.7736$. Mechanism decomposition shows the two mechanisms contribute differently across index families and should be dosed separately, yielding a safety boundary $0<c\leq0.3$, a plateau $\omega\in[0.5,12]$ and a near-critical plateau $\sigma\in[0.7,0.95]\sigma^{*}$. Two negative results bound the claim: under a degree-corrected benchmark the gains over local baselines vanish and reverse in sign, and the AUC gains do not translate into top-$L$ hit rate.

\end{abstract}

\section{Introduction}

Complex networks appear widely in social, biological, transportation and information systems. Link prediction aims at inferring missing or future links from the observed connection structure of a network, and is used in friend recommendation, protein-interaction prediction, academic-collaboration mining and collaborative filtering. Liben-Nowell and Kleinberg~\cite{A8} established link prediction as an independent research problem and laid down the basic paradigm of ``computing a similarity score for each node pair''. Over the following two decades, link prediction methods gradually formed a three-tier methodological system dominated by local, global and quasi-local methods~\cite{A8,A14,A16}.

These three tiers, however, trade off structural vision against computational cost at both ends. Local methods rely only on first-order neighborhood information; representative indices include Common Neighbors (CN)~\cite{A8}, the Jaccard coefficient (JC)~\cite{A12}, the Adamic--Adar index (AA)~\cite{A9}, the Resource Allocation index (RA)~\cite{A10}, Preferential Attachment (PA)~\cite{A11}, the Hub Promoted Index (HPI)~\cite{A26} and the S{\o}rensen index (SO)~\cite{A27}. They are cheap and interpretable, but assign a constant zero score to node pairs without common neighbors and therefore perform poorly on sparse and tree-like networks. Global methods exploit whole-network topology, such as the Katz index~\cite{A13}, Random Walk with Restart (RWR)~\cite{A14} and SimRank~\cite{A15}; they are more accurate, but their matrix inversion or iterative convergence costs up to $O(n^3)$, which prohibits application to very large networks. Quasi-local methods (LRW~\cite{A16}, LPI~\cite{A17}, CNC~\cite{A19}, CND~\cite{A18}, etc.) strike a compromise between accuracy and cost, yet remain limited by their strong dependence on common-neighbor information. Subsequent studies further enriched the methodology around structural heterogeneity and ensemble strategies: Local Naive Bayes (LNB)~\cite{A20} quantifies the promoting or inhibiting role of each common neighbor; mutual information (MI)~\cite{A25} measures statistical dependence between nodes; Shang et al.~\cite{A21} proposed heterogeneity-based predictors for sparse, tree-like networks; the local-community paradigm (CAR)~\cite{A23} incorporates community information; CN2D~\cite{A22} and SAM~\cite{A24} complement local information from the perspectives of neighbor degree distribution and directionality, respectively. A systematic survey of the methodology and recent progress can be found in~\cite{A28}; critical studies of evaluation pitfalls in GNN-based link prediction also show that, under fair tuning, well-trained simple heuristic baselines remain highly competitive; purely topological similarity information thus remains irreplaceable in attribute-free, resource-constrained scenarios~\cite{A30}.

The core question is how to break the structural-vision limitation of local methods while preserving their low complexity. One effective route is the \textbf{fusion of node centrality indices with local similarity indices}. Node centrality quantifies the relative importance of a node in the network and can provide completion scores for node pairs without common neighbors; local similarity captures the fine structure of node neighborhoods and retains sensitivity to the distribution of common neighbors. Charikhi~\cite{A3} incorporated PageRank scores into local indices with multiplicative weights and reported stable improvements on 11 network datasets; this is the representative work of the fusion paradigm. A closer look, however, reveals three open problems. First, its weight coefficients must be tuned index by index, lacking a unified form applicable to all indices, and the statistical significance and overfitting risk of per-index tuning have never been discussed. Second, existing works adopt heterogeneous evaluation protocols (the construction of negative samples, the fold partition of cross-validation), and some protocols even risk filtering candidate negative samples by the very scores being evaluated; whether the reported improvements survive a fair, exhaustive-comparison protocol lacks independent verification. Third, the dose design of the two mechanisms (``modulation'' and ``completion'') in the fusion rule, i.e., the admissible ranges of the weight and the completion coefficient, relies entirely on experience, with no quantitative design principles to follow.

The idea of fusing centrality with local similarity has several neighboring implementations in the link prediction literature, and the position of the present work within this line of research needs to be clarified first. The CCPA algorithm of Ahmad et al.~\cite{A19} parametrically weights the contribution of common neighbors by node centrality and is the most direct representative of ``centrality-enhanced local indices''; Wahid-Ul-Ashraf et al.~\cite{A36} borrowed the law of gravitation, using node centrality as ``mass'' and pairwise distance (or inverse similarity) as ``distance'' to construct a gravity-style score, coupling global importance explicitly into the connection propensity of a node pair; Nandini et al.~\cite{A37} modified common-neighbor indices using the average centrality of the network. Two further variants belong to the same family and are closer to the present work in spirit: Liu et al.~\cite{A50} weight each common neighbor by its own centrality (degree, closeness or betweenness) and add a weak-tie term, and Li et al.~\cite{A49} extend CCPA with local-path and closeness-centrality terms. All of these modulate or reweight the contribution of individual common neighbors, whereas our rule never touches the common-neighbor term itself: it multiplies or completes the aggregated local score with a node-level centrality product, so the object being corrected is the local index as a whole rather than its summands. Several independent concurrent works have appeared in the last two years: He et al.~\cite{A40} constructed multi-order popularity and similarity predictors based on graphlet-orbit degrees and systematically demonstrated, on 550 real networks, the complementarity of ``popularity'' and ``similarity'' mechanisms in network formation; Saifi et al.~\cite{A39} proposed a family of ``popularity degree'' indices to complete common-neighbor information; Chen et al.~\cite{A38} combined HITS centrality with biased random walks for directed networks; Nandini et al.~\cite{A41} generalized multiple centralities to hypernetwork scenarios. We do not aim to propose yet another fusion instance; rather, we answer three foundational questions of the paradigm itself: \textbf{Under what evaluation protocol is a fusion improvement credible? What properties must a centrality index satisfy to fuse effectively? How should the fusion parameters be chosen?}

The DomiRank centrality~\cite{A6}, recently proposed by Engsig et al., is a node-dominance measure that balances local neighborhood and mesoscale topological information through competition dynamics, and possesses the closed-form expression $\bm{\Gamma}^{*}=\theta\sigma(\sigma A+I)^{-1}A\bm{1}$; its tunable parameter $\sigma$ provides a controlled setting for studying ``which property of a centrality index determines the fusion effect''. A systematic study of the mathematical properties of this centrality, namely its behaviour under graph automorphisms, its spectral entropy, and its invariance under graph transformations, is given in a companion manuscript~\cite{A47}. Beyond PageRank and DomiRank, k-shell decomposition~\cite{A32}, HITS~\cite{A33}, betweenness centrality~\cite{A34}, eigenvector centrality~\cite{A35} and GNN-based learned centralities~\cite{A29,A31} can also play the role of ``global completion'' within the same framework.

We carry out our study within a unified piecewise ``centrality$\,\times\,$local similarity'' fusion framework. The main contributions are as follows, aligned with the two halves of the title: a fair-protocol reassessment (1)--(2) and parameter design principles (3)--(4). (1) We give the general form of the fusion rule (multiplicative modulation $+$ small-dose completion) and show, through a scale analysis and a diagnostic of information redundancy, that the \textbf{scale calibration and information content of the centrality product} jointly determine whether the modulation mechanism works: the PageRank product is of order $O(n^{-2})$, so its multiplicative factor is a near-identity map, and on undirected networks PageRank is approximately a monotone function of node degree (rank correlation above $0.89$), so its product is highly redundant with the preferential-attachment index; this predicts theoretically the failure boundary of PageRank fusion. (2) \textbf{Fair-protocol reassessment}. Under a fair protocol with exhaustive negative-sample comparison and no score-based filtering whatsoever, we independently reassess PageRank fusion: for six of the seven local indices (SO being a monotone transform of JC, PA a degree-based reference) the fusion gains lack consistent cross-network statistical significance, clarifying the applicability boundary of existing conclusions of the paradigm; through three control arms (order-preserving scaling, min-max normalization and a normalized degree product, Section~\ref{sec:ctrl}) we further show that scale calibration is a necessary but not a sufficient condition, and that non-redundancy with the local information is what actually decides whether the modulation mechanism takes effect; the single exception is the near-tree-like wiki-Vote network, where a uniform gain of about $+0.036$ (the largest single-dataset gain in this paper) arises entirely from the completion mechanism, and is confirmed by an internal control, namely that the gain of the PA index on that network is merely $+5\times10^{-6}$, consistent with the testable prediction of the scale analysis. (3) \textbf{Unified parameters and baseline comparison}. We introduce the DomiRank centrality and show that, after min-max normalization, under \textbf{unified parameters} ($\omega=5$, $c=0.1$) the normalized fused indices DR-JC/HPI/SO significantly outperform their local baselines already at the dataset level (8 independent networks, Wilcoxon and sign tests $p=0.027$ and $0.035$), while the remaining four indices obtain direction-consistent gains that are significant at the fold level ($p\leq3.3\times10^{-3}$, BH-corrected $q\leq4.9\times10^{-3}$) but do not reach dataset-level significance; in particular, DR-RA outperforms nine of the ten comparison methods (Katz, LNB, CN2D, CNC, CND, SimRank, CCPA, Gravity and CNPop) and is comparable to RWR; a $10\times5$ repeated cross-validation confirms that these conclusions are robust to fold-partition randomness. (4) \textbf{Parameter design principles}. We provide quantitative design principles for the fusion paradigm: a safety boundary $0<c\leq0.3$ for the completion coefficient, a wide plateau $\omega\in[0.5,12]$ for the modulation weight, and a near-critical plateau $\sigma\in[0.7,0.95]\sigma^{*}$ for the DomiRank competition intensity; mechanism-decomposition experiments further clarify the differentiated contributions of completion and modulation to different families of local indices: the completion gain is jointly determined by the fraction of zero-score node pairs and the standalone predictive power of the centrality product, while the effectiveness of modulation is determined by the dynamic range of the centrality product, so the doses of the two mechanisms should be designed separately.

\section{Preliminaries}

\subsection{Node centrality indices: PageRank and DomiRank}

The two centrality indices considered in this paper characterize the global importance of nodes from the perspectives of ``random walk'' and ``local competition'', respectively, and are two representative instantiations of the fusion paradigm.

\subsubsection{PageRank (random-walk view)}

The PageRank algorithm, proposed by Page and Brin~\cite{A1}, simulates the random-walk behavior of a user on the network and quantifies the relative importance of nodes via directed link relations. The algorithm introduces a damping factor $d$ (usually $d=0.85$): the walker follows an outgoing edge of the current node with probability $d$ and jumps to an arbitrary node with probability $1-d$, which guarantees convergence of the iteration. The PageRank value of node $u$ satisfies the iterative relation
\begin{equation}
PR(u) = \frac{1-d}{n} + d \sum_{v\in B(u)} \frac{PR(v)}{c(v)},
\end{equation}
where $B(u)$ is the set of nodes pointing to $u$, $c(v)$ is the out-degree of node $v$, and $n$ is the number of nodes. On the undirected networks of Section~3.3 each edge is read as two opposite arcs, so $B(u)=\Gamma_u$ and $c(v)=|\Gamma_v|$ and the transition matrix is symmetric. PageRank is widely used in graph ranking, web search and influence analysis in social networks. For the present paper, a key property of PageRank is that its scores form an $n$-dimensional probability distribution ($\sum_u PR(u)=1$) and are therefore typically of order $O(1/n)$; Section~4 will show that this scale property directly determines the ceiling of its fusion effect.

\subsubsection{DomiRank (competition-dynamics view)}

DomiRank, proposed by Engsig et al.~\cite{A6}, quantifies the dominance of a node within its neighborhood. The node-dominance vector $\bm{\Gamma}(t)$ obeys the competition-dynamics equation
\begin{equation}
\frac{\mathrm{d}\bm{\Gamma}(t)}{\mathrm{d}t} = \alpha A \bigl(\bm{1}_{N\times 1} - \bm{\Gamma}(t)\bigr) - \beta \bm{\Gamma}(t),
\end{equation}
where $A$ is the adjacency matrix of the network, $\alpha$ the competition intensity, $\beta$ the natural decay rate, and $\bm{1}$ the all-ones vector. Letting $\sigma = \alpha/\beta \in \mathbb{R}^{+}$ and assuming $\det(\sigma A + I) \neq 0$, the steady-state solution has the closed form
\begin{equation}
\bm{\Gamma}^{*} = \theta\sigma(\sigma A + I)^{-1} A \bm{1}_{N\times 1},
\end{equation}
where $\theta$ is the upper bound of dominance strength (usually $\theta=1$) and $I$ is the identity matrix. The parameter $\sigma$ balances local and mesoscale topological information: for small $\sigma$, node scores mainly reflect local features such as the number of direct neighbors; as $\sigma$ increases, the relative position of a node in the whole-network competition is emphasized. Following the convention of Engsig et al., we express $\sigma$ as a relative proportion of the critical value $\sigma^{*}=-1/\lambda_{\min}(A)$ ($\lambda_{\min}(A)$ being the smallest eigenvalue of the adjacency matrix), i.e., the actual computation uses $\sigma_{\mathrm{eff}}=\sigma\cdot\sigma^{*}$ with $\sigma\in(0,1]$ corresponding to the effective domain. The closed-form solution $\bm{\Gamma}^{*}$ may contain negative values; throughout this paper we min-max normalize it to $[0,1]$ before fusion:
\begin{equation}
\widetilde{\bm{\Gamma}} = \frac{\bm{\Gamma}^{*} - \min_i \Gamma^{*}_i}{\max_i \Gamma^{*}_i - \min_i \Gamma^{*}_i}.
\label{eq:minmax}
\end{equation}
Sections~4 and~6 will show that this normalization is not an implementation detail but a design element that determines the success of fusion.

\subsubsection{Duality of the two centralities and their fusion roles}

Although PageRank and DomiRank differ significantly in dynamical origin, parameter semantics and computational complexity, their roles in the fusion paradigm coincide in two respects: both map a node to a scalar representing its global importance and provide a global score for a node pair in the form $\mathrm{Centrality}(x)\cdot \mathrm{Centrality}(y)$. Their difference in fusion effect (Sections~4 and~5) constitutes a natural contrast: the former has a product scale of $O(n^{-2})$, while the latter, after normalization, has a scale of $O(1)$, a natural experiment for the claim that ``scale determines fusion effectiveness''.

\subsection{The fusion paradigm: general framework and design questions}

Let $v_x$, $v_y$ be a pair of unconnected nodes in the network, and $\Gamma_x$ the neighbor set of $v_x$. A local similarity index $S^{MD}(x,y)$ depends on the common neighbors $\Gamma_x \cap \Gamma_y$ and scores zero for node pairs without common neighbors. Let $C(\cdot)$ be any centrality index mapping a node to a scalar of global importance, and define the centrality similarity
\begin{equation}
S^{C}(x,y) = C(x) \cdot C(y).
\end{equation}
Intuitively, $S^{C}(x,y)$ characterizes the product of the individual ``importance'' of the two nodes and can be regarded as a potential driving force for connection between them; it remains nonzero even in the absence of common neighbors and can therefore serve as a completion of the local score. We adopt the following unified piecewise fusion rule:
\begin{equation}
S_{\mathrm{fused}}(x,y) =
\begin{cases}
S^{MD}(x,y)\cdot \bigl(1 + \omega \cdot S^{C}(x,y)\bigr), & |\Gamma_x \cap \Gamma_y| \neq 0 \text{ or } MD = PA, \\[4pt]
c \cdot S^{C}(x,y), & |\Gamma_x \cap \Gamma_y| = 0 \text{ and } MD \neq PA,
\end{cases}
\label{eq:fusion}
\end{equation}
where $\omega\geq 0$ is the modulation weight and $c\geq 0$ the completion coefficient ($c=0$ degenerates to pure modulation without completion). This form contains two mechanisms: \textbf{multiplicative modulation}, an order-preserving amplification of the local score by centrality when local information is sufficient; and \textbf{small-dose completion}, filling the zero scores with the centrality product when local information is absent. The Charikhi method~\cite{A3} is the special case of Eq.~\eqref{eq:fusion} with $C=\mathrm{PR}$, $c=5$ and per-index values of $\omega$.

Equation~\eqref{eq:fusion} reduces fusion design to three questions: (i) In what range should the magnitude of $S^{C}$ lie so that the multiplicative factor $1+\omega S^{C}$ produces substantive modulation while the completion term $c\cdot S^{C}$ does not overwhelm the ordering of local scores? (ii) How should $(\omega,c)$ be chosen uniformly? (iii) When evaluating fusion effects, how should negative samples be constructed to avoid circular reasoning? Section~3 first establishes a fair evaluation protocol; Section~4 uses PageRank to give a scale analysis of question (i) together with reassessment evidence; Section~5 uses DomiRank to answer question (ii) and provides the complete parameter design space.

Regarding computational complexity, solving a local index $S^{MD}$ costs $O(vk)$ ($v$ being the number of nodes and $k$ the maximum degree); the iterative solution of PageRank costs $O(d\cdot|E|)$ ($d$ being the number of iterations); the closed-form solution of DomiRank requires solving one linear system with $(\sigma A + I)$, whose direct dense inversion costs $O(v^3)$, but Engsig et al.~\cite{A6} have provided an efficient parallelizable implementation that makes it applicable to massive networks. In this paper, networks with $|V|>800$ are solved by sparse LU factorization (verified digit-by-digit against the dense solution), and a single solve on the largest network (7066 nodes) takes only about 5 seconds. The overall computational complexity of the fusion framework is dominated by the chosen centrality index.

\section{Similarity indices and experimental design}

\subsection{Local similarity indices}

We select seven classical local similarity indices as the base measures of the fusion framework. Common Neighbors (CN)~\cite{A8} directly counts the common neighbors of two nodes:
\begin{equation}
S^{CN}(x,y) = |\Gamma_x \cap \Gamma_y|.
\end{equation}
The Adamic--Adar index (AA)~\cite{A9} weights each common neighbor by the inverse logarithm of its degree:
\begin{equation}
S^{AA}(x,y) = \sum_{z\in \Gamma_x \cap \Gamma_y} \frac{1}{\log |\Gamma_z|}.
\end{equation}
The Resource Allocation index (RA)~\cite{A10} weights each common neighbor by the inverse of its degree:
\begin{equation}
S^{RA}(x,y) = \sum_{z\in \Gamma_x \cap \Gamma_y} \frac{1}{|\Gamma_z|}.
\end{equation}
The Jaccard index (JC)~\cite{A12} is the normalized form of the common-neighbor fraction:
\begin{equation}
S^{JC}(x,y) = \frac{|\Gamma_x \cap \Gamma_y|}{|\Gamma_x \cup \Gamma_y|}.
\end{equation}
The Hub Promoted Index (HPI)~\cite{A26} has a denominator determined by the lower-degree node:
\begin{equation}
S^{HPI}(x,y) = \frac{|\Gamma_x \cap \Gamma_y|}{\min(|\Gamma_x|,|\Gamma_y|)}.
\end{equation}
The S{\o}rensen index (SO)~\cite{A27} is a monotone transform of JC: letting $r = |\Gamma_x\cap\Gamma_y|/(|\Gamma_x|+|\Gamma_y|)$, one has $SO=2r$, and $r/(1-r) = |\Gamma_x\cap\Gamma_y|/(|\Gamma_x|+|\Gamma_y|-|\Gamma_x\cap\Gamma_y|) = |\Gamma_x\cap\Gamma_y|/|\Gamma_x\cup\Gamma_y| = JC$, an identity valid for arbitrary node pairs. Since both indices are functions of the same two quantities $|\Gamma_x\cap\Gamma_y|$ and $|\Gamma_x\cup\Gamma_y|$, their rank-based AUCs are necessarily identical:
\begin{equation}
S^{SO}(x,y) = \frac{2 \times |\Gamma_x \cap \Gamma_y|}{|\Gamma_x| + |\Gamma_y|}.
\end{equation}
Preferential Attachment (PA)~\cite{A11} measures connection propensity by the product of the two degrees and does not itself depend on common neighbors:
\begin{equation}
S^{PA}(x,y) = |\Gamma_x| \times |\Gamma_y|.
\end{equation}

\subsection{Global and advanced baseline indices}

To examine the relative position of the fused indices within the broader family of methods, we select ten comparison baselines, which we introduce in two groups: first seven, of which three are global methods and four are advanced similarity indices, and then three further centrality-based ones (CCPA, Gravity and CNPop) at the end of this subsection. The scope of the comparison is stated explicitly: all methods compared here are attribute-free and non-learning structural methods, for which no training split and no node features are required. Learned and embedding-based approaches are not part of our comparison family; a fair-protocol evaluation of graph neural network link predictors reports that well-trained simple heuristic baselines remain highly competitive under equal tuning~\cite{A30}, so the conclusions of this paper do not rest on a direct comparison against that family. The Katz index~\cite{A13} accumulates all paths between two nodes with length-decayed weights:
\begin{equation}
S^{KI}(x,y) = \sum_{l=1}^{\infty} \alpha^l \cdot (A^l)_{xy} = \bigl((I-\alpha A)^{-1} - I\bigr)_{xy},
\end{equation}
where $\alpha$ is the decay factor. The RWR index~\cite{A14} describes a random walk starting from node $v_x$:
\begin{equation}
\pi_x(t+1) = c \cdot \bm{P}^{\mathsf{T}} \pi_x(t) + (1-c) e_x, \quad S^{RWR}(x,y) = \pi_{xy} + \pi_{yx}.
\end{equation}
The SimRank index~\cite{A15} is based on the recursive assumption that ``similar nodes have similar neighbors'':
\begin{equation}
S^{SR}(x,y) =
\begin{cases}
1, & x = y,\\
\displaystyle\frac{r}{|\mathrm{In}(x)||\mathrm{In}(y)|} \sum_{i=1}^{|\mathrm{In}(x)|}\sum_{j=1}^{|\mathrm{In}(y)|} S^{SR}(\mathrm{In}_i(x), \mathrm{In}_j(y)), & x\neq y,
\end{cases}
\end{equation}
where $\mathrm{In}(x)$ is the in-neighbor set of $v_x$ and $r$ is the damping factor. The LNB index~\cite{A20} quantifies the role of common neighbors within a naive-Bayes framework:
\begin{equation}
S^{LNB}(x,y) = \sum_{z\in \Gamma_x \cap \Gamma_y} \log\bigl(o \cdot R_z\bigr),
\end{equation}
where $o=\frac{|V|(|V|-1)}{2|E|}-1$ is the prior odds of disconnection over connection, and $R_z=\frac{C_z+1}{D_z+1}$ ($C_z$ and $D_z$ being the numbers of links and non-links among the neighbors of $v_z$, respectively); this is the logarithmic form of the original model score $\prod_{z\in\Gamma_x\cap\Gamma_y} o\,R_z$, in which a pair-independent constant factor is omitted and does not affect the ranking. The CN2D index~\cite{A22} introduces a correction based on the degree distribution of common neighbors:
\begin{equation}
S^{CN2D}(x,y) = |\Gamma_x \cap \Gamma_y| + \frac{\alpha}{\max(|\Gamma_x|,|\Gamma_y|)} \sum_{z\in \Gamma_x\cap\Gamma_y} |\Gamma_z|.
\end{equation}
The CNC and CND indices~\cite{A19,A18} respectively incorporate a closeness-type distance correction and a dynamic switching strategy based on the presence of common neighbors (the idea of weighting common neighbors by centrality also appears in the multi-network setting of Nasiri et al.~\cite{A43}):
\begin{equation}
S^{CNC}(x,y) = \alpha\, \frac{|\Gamma_x \cap \Gamma_y|}{|\Gamma_x \cup \Gamma_y|} + (1-\alpha)\, \frac{1}{\mathrm{dist}(x,y)},
\label{eq:cnc}
\end{equation}
\begin{equation}
S^{CND}(x,y) =
\begin{cases}
\frac{|\Gamma_x \cap \Gamma_y| + 1}{l}, & |\Gamma_x \cap \Gamma_y| \neq 0,\\[4pt]
\displaystyle\frac{k}{\mathrm{dist}(x,y)}, & \text{otherwise},
\end{cases}
\label{eq:cnd}
\end{equation}
where $l=3$ and $k=2$ are normalization constants. In Eq.~\eqref{eq:cnd} the distance-based branch is active only for pairs without common neighbors, which is exactly the branch that this paper's fusion rule targets (Section~2.2); the two distance-aware baselines compared here are therefore CNC in Eq.~\eqref{eq:cnc} and CCPA in Eq.~\eqref{eq:ccpa}, and all three are kept in their reported forms without tuning. In the CNC score the common-neighbor count is normalized by the union size (i.e., the Jaccard coefficient); this normalization is essential here, because the raw count form $|{\Gamma_x\cap\Gamma_y}|$ is structurally order-equivalent to CCPA below (a node pair with a common neighbor is always at distance two, so the distance term is constant within that group, and pairs without common neighbors are ordered by distance alone), which would make the two baselines numerically identical.

To align with recent centrality-based fusion works, we additionally take three advanced baselines: CCPA~\cite{A19}, which compensates the common-neighbor count with a pair-level harmonic-closeness-type centrality term,
\begin{equation}
S^{CCPA}(x,y) = 0.5\,|\Gamma_x \cap \Gamma_y| + 0.5\,\frac{n}{\mathrm{dist}(x,y)};
\label{eq:ccpa}
\end{equation}
the term $n/\mathrm{dist}(x,y)$ is precisely what Ahmad et al.~\cite{A19} call the ``centrality'' term of a node pair (a harmonic-closeness-type quantity), and the formula above is the original CCPA functional form (see also the reference implementation \texttt{common\_neighbor\_centrality} in NetworkX), with the weight set to $\alpha=0.5$ instead of the $0.8$ recommended in~\cite{A19}. The reason for this choice is disambiguation from the CNC baseline above: at $\alpha=0.8$ and with raw common-neighbor counts, CCPA coincides with the raw-count form of CNC (see the order-equivalence remark above), so the two baselines would be numerically identical; $\alpha=0.5$ keeps them distinct while preserving the original functional form. Gravity~\cite{A36}, which borrows the law of gravitation with node degree as ``mass'':
\begin{equation}
S^{G}(x,y) = \frac{k_x k_y}{\mathrm{dist}(x,y)^2};
\end{equation}
(here $\mathrm{dist}(x,y)$ denotes the graph distance, the same notation as in Eqs.~\eqref{eq:cnc}--\eqref{eq:ccpa}; it is unrelated to the damping factor $d$ of Eq.~(1)), and CN-Popularity (CNPop)~\cite{A42}, which modulates the common-neighbor count by the sum of degrees:
\begin{equation}
S^{CNPop}(x,y) = |\Gamma_x \cap \Gamma_y| \cdot (k_x + k_y).
\end{equation}

\subsection{Datasets}

We adopt eight real-world network datasets with diverse topological characteristics (all treated as unweighted and undirected, with self-loops removed and the largest connected component taken), covering political books, neural, academic collaboration, music social, e-mail, online social and Wikipedia voting domains; their basic properties are listed in Table~\ref{tab:data}. The node sizes span about two orders of magnitude, and the average degree and the clustering coefficient span roughly a factor of nine and a factor of five respectively (Table~\ref{tab:data}). These spans in scale, density and clustering structure are wide enough to test how the fusion strategy adapts to different topologies. Among them, three networks (ca-GrQc, facebook\_combined and wiki-Vote) have more than $10^3$ nodes and email-Eu-core has 986 nodes (largest connected component), just below $10^3$; all four are taken from the SNAP collection~\cite{A44} and are used to examine the performance of the fused indices on larger networks.

\begin{table}[!t]
	\centering
	\caption{Basic properties of the 8 real-world network datasets}
	\small
	\begin{tabular}{lccccc}
		\toprule
		Dataset & Type & $|V|$ & $|E|$ & $\langle k\rangle$ & $C$\\
		\midrule
		BUP & Political books & 105 & 441 & 8.40 & 0.488\\
		CEG & Neural & 297 & 2148 & 14.46 & 0.292\\
		NS & Collaboration & 379 & 914 & 4.82 & 0.741\\
		Jazz & Jazz musicians & 198 & 2742 & 27.70 & 0.617\\
		Email & E-mail & 986 & 16064 & 32.58 & 0.407\\
		FB & Social & 4039 & 88234 & 43.69 & 0.606\\
		GrQc & Collaboration & 4158 & 13422 & 6.46 & 0.557\\
		Wiki & Politics & 7066 & 100736 & 28.51 & 0.142\\
		\bottomrule
	\end{tabular}
	\label{tab:data}
\end{table}

\subsection{Evaluation methodology}

We evaluate all indices with five-fold cross-validation~\cite{A4}: the observed edges of a network are randomly and evenly divided into five subsets; each subset is taken in turn as the test set (probe edges) and the remaining four as the training set; the average performance over the five rounds is taken as the final result, and the per-fold AUCs are recorded to estimate fluctuations and to conduct statistical tests. The evaluation metric is AUC~\cite{A7}, the standard measure in link prediction:
\begin{equation}
\mathrm{AUC} = \frac{n_1 + 0.5 n_2}{n},
\end{equation}
where $n$ is the number of independent comparisons, $n_1$ the number of times a missing link scores higher than a nonexistent link, and $n_2$ the number of ties. $\mathrm{AUC}=0.5$ corresponds to random guessing, and values closer to $1$ indicate higher prediction accuracy.

To examine the precision at the top of the prediction list, we also report Precision@$L$: with $L$ set to the number of test edges per fold, all candidate pairs (non-links of the training graph together with the test edges) are ranked by score in descending order, and Precision@$L$ is the fraction of test edges among the top-$L$ pairs. This metric is sensitive to the ranking quality at the top and complements AUC, which reflects only the global ordering and does not distinguish head hits.

We fix the \textbf{fairness} of the evaluation protocol explicitly as follows: for each fold, the negative samples are taken as \textbf{all} non-links of the training graph (probe edges excluded), and an \textbf{exhaustive} pairwise comparison against the probe edges is performed, without any subsampling, and without any filtering or screening of candidate negative samples by the scores of any method under evaluation. This choice is based on two considerations: first, as a pairwise ranking probability, AUC is unbiased under uniform subsampling of negative samples, and exhaustive comparison eliminates sampling variance; second, any screening of negative samples based on the evaluated scores constructs an evaluation benchmark favorable to that method, forming circular reasoning whose conclusions are not credible. The split is even by construction, so the probe-edge ratio is fixed at \(20\%\) of the observed edges in every fold and on every network; no dataset-dependent hold-out fraction enters the protocol. All implementations in this paper (index computation, fold partition, AUC computation) use fixed random seeds ($1000+k$, $k$ being the fold index), and all results are directly reproducible. The seed rule is uniform in every experiment reported here except the repeated cross-validation, which is deliberately independent: the parameter grid search of Section~5.2, the degree-corrected arms of Section~5.7 (each arm shares the fold partition of arm~$A$ within the same pass), the mechanism decomposition of Section~5.6 and the leave-one-network-out validation all regenerate the edge permutation from $1000+k$ on the network in question, so two experiments on the same network and fold index are partition-identical. The $10\times5$ repeated cross-validation of Section~5.3 instead uses $5000+100r+k$ for repetition $r\in\{0,\dots,9\}$ and fold $k\in\{0,\dots,4\}$, a disjoint seed family chosen so that its partitions are not a resampling of the main ones; this is why that experiment provides genuine independent evidence of robustness to fold-partition randomness, whereas the other sections reuse the main partition by design, so that all parameter choices are compared on identical data.

Regarding statistical significance, we conduct one-sided Wilcoxon signed-rank tests (significance level $\alpha=0.05$) for paired indices (DR-MD against local baselines, DR-MD against PR-MD, and DR-RA against each comparison method) on the 40 fold-level paired observations of a single index (8 datasets $\times$ 5 folds). The five fold observations of the same dataset share the training-graph structure and are not fully independent, so the fold-level $p$-values serve only as suggestive evidence; we therefore supplement them with two more robust classes of tests: (i) \textbf{dataset-level tests}: taking the eight networks (each represented by its five-fold mean) as paired units, we perform Wilcoxon signed-rank tests and sign tests of DR-MD against the local baselines; the sample size is small but the test units are truly independent; (ii) \textbf{multiple-comparison correction}: Benjamini--Hochberg (BH) FDR correction is applied uniformly to the 31-test BH family of this paper, which comprises the 31 fold-level paired tests (7 PR-reassessment tests, 7 DR-versus-local tests, 7 DR-versus-PR tests and 10 tests of DR-RA against global/advanced baselines), and the corrected $q$-values are reported to control the false discovery rate. In addition, since JC and SO are monotone transforms of each other (Section~3.1), their fold-level observations are not independent, so we never pool observations across indices: all tests are carried out strictly within a single index, and the two near-duplicate JC/SO tests are nevertheless retained in the 31-test BH family because they enter different comparison groups (PR reassessment, DR-versus-local, DR-versus-PR) and dropping either one would alter the family composition asymmetrically. The resulting positive dependence concerns two of the 31 tests only, a setting in which the BH procedure remains valid to a good approximation. Finally, the one-sided alternative is fixed \emph{a priori} by the design of the fusion rule: every test asks whether the prespecified direction (the fused or tuned variant is better) holds, not whether any difference exists.
Because the dataset-level family contains only eight independent units, the resolution of such a test has to be spelled out. For the exact one-sided Wilcoxon signed-rank test on $n=8$ units with no zero differences, the attainable $p$-values are $0.0039$ for $8/8$ wins, $[0.0078, 0.0977]$ for $7/8$, $[0.0195, 0.3711]$ for $6/8$, $[0.0547, 0.6797]$ for $5/8$ and $[0.1562, 0.8750]$ for $4/8$; the corresponding exact one-sided sign test gives $0.0352$, $0.1445$, $0.3633$ and $0.6367$ for $7/8$, $6/8$, $5/8$ and $4/8$.

Two consequences are used in what follows. (i) The resolution floor is $p=0.0039$, so the dataset-level test is not vacuous: the significant results on DR-JC/HPI/SO ($p=0.027$) and the non-significant ones on CN/AA/RA/PA ($p$ between $0.055$ and $0.12$, Section~5.3) both lie inside the attainable range for their respective win counts, and the latter are therefore genuine non-significance at $6/8$ rather than an artefact of insufficient resolution. (ii) An $n=8$ dataset-level test cannot certify an effect that a minority of networks contradicts: significance at this level requires that at least six of the eight networks agree, so Section~5.2 backs the dataset-level test with the $10\times5$ repeated cross-validation of Section~5.3. Where the signed-rank distribution is symmetric the two-sided $p$-values are twice the reported ones; we flag explicitly the single borderline case affected by this choice, namely the dataset-level significance of DR-JC/HPI/SO ($p=0.027$ one-sided, $\approx 0.055$ two-sided), and no other conclusion of the paper depends on the one- versus two-sided distinction. (iii) The enumeration above is stated for the eight dataset-level units of Table~\ref{tab:wilcox}, and it does not transfer to the forty fold-level pairs pooled in Table~\ref{tab:degcorr_gain}. Those pairs have an $n$ eight times larger, so their attainable $p$-values are correspondingly finer; for those comparisons the quantity to read is the positive-pair count (Pos./40) reported there, not this enumeration. (iv) The intervals above presuppose no zero differences, since the enumeration runs over the $2^{n}$ sign patterns of the signed-rank statistic. We verified this precondition rather than assuming it: in the experiment records underlying both families, namely arms~$A$ and~$B$ of Section~\ref{sec:degcorr}, seven indices by forty pairs each, the number of zero differences is $0$ in every one of the fourteen cells, so the effective sample size equals the nominal one throughout and no reported $p$ value is affected by the zero-handling convention of the Wilcoxon implementation.

\section{A fair-protocol reassessment of PageRank fusion}

\subsection{Method definition}

Taking $C=\mathrm{PR}$ in Eq.~\eqref{eq:fusion} yields the Charikhi-type PageRank fused indices~\cite{A3}:
\begin{equation}
S_{PR}^{MD}(x,y) =
\begin{cases}
S^{MD}(x,y) \cdot \bigl(1 + \omega_{MD} \cdot S^{PR}(x,y)\bigr), & |\Gamma_x \cap \Gamma_y| \neq 0 \text{ or } MD = PA, \\[6pt]
5 \cdot S^{PR}(x,y), & |\Gamma_x \cap \Gamma_y| = 0 \text{ and } MD \neq PA,
\end{cases}
\label{eq:prfusion}
\end{equation}
where $S^{PR}(x,y)=PR(x)\cdot PR(y)$; the weight coefficients follow the experimental values reported in the literature, $\omega_{CN}=1.8$, $\omega_{AA}=1.8$, $\omega_{RA}=0.8$, $\omega_{JC}=1.2$, $\omega_{HPI}=1.5$, $\omega_{SO}=1.2$, $\omega_{PA}=2$, and the completion coefficient is $5$.

\subsection{Scale analysis: why the multiplicative factor is nearly identity}

PageRank scores form an $n$-dimensional probability distribution with $\sum_u PR(u)=1$ and a typical magnitude of $O(1/n)$, hence
\begin{equation}
S^{PR}(x,y) = PR(x)\cdot PR(y) \approx O(n^{-2}).
\end{equation}
For the experimental networks of this paper ($105\leq n\leq 7066$), $S^{PR}$ ranges from about $2\times 10^{-8}$ to $9\times 10^{-5}$ (the extremes of the node-product over candidate pairs, whereas the median magnitude over candidate pairs reported in Section~4.4 is of order $10^{-5}$, two to three orders of magnitude smaller). The multiplicative factor therefore satisfies
\begin{equation}
1 + \omega_{MD}\cdot S^{PR}(x,y) \approx 1 + O(10^{-4}),
\end{equation}
which, for $\omega_{MD}\leq 2$, changes each local score by no more than $0.02\%$ relatively, so that, except for nearly tied node pairs, this factor \textbf{does not change the ordering of local scores}. That is, the first branch of Eq.~\eqref{eq:prfusion} is mathematically a near-identity map, and tuning $\omega_{MD}$ index by index amounts to turning an ineffective knob; the fusion effect can only come from the completion term $5\cdot S^{PR}$ of the second branch, whose magnitude ($\sim 10^{-4}$) is far smaller than any positive local score (e.g., $CN\geq 1$) and whose effect is limited to inducing an $S^{PR}$-determined order among the zero-score node pairs. Two directly testable corollaries follow. First, $\omega_{MD}$ is indeed a dead knob: with $c=0$ (completion switched off, pure modulation) and $\omega$ swept over $\{0.5,1,1.8,3,5,8,12\}$ on the full parameter grid of the four original datasets (PR fusion, all seven indices, $4\times5=20$ fold-level values per cell), the mean $\Delta$AUC of the seven indices is identically $+0.00068$ at every value of $\omega$, with the per-index values confined to $+0.00007\sim +0.00164$ and matching to five decimal places across the whole sweep, so that enlarging $\omega$ by a factor of 24 changes nothing; by contrast, DR fusion on the same grid and the same seven indices varies monotonically with $\omega$, its seven-index mean gain rising from $+0.00174$ at $\omega=0.5$ to $+0.00430$ at $\omega=12$. Both scans use the same four datasets, the same fold partitions and the same local baselines; they differ only in the centrality fused (PR versus DR), and the two families should not be compared term by term, since the dead-knob phenomenon of Section~4.2 is specific to the unnormalized PR product analysed above. Second, $S^{PR}$ is highly redundant with local information: on the four networks the Spearman rank correlation between the node PageRank value and its degree is $0.976$ (BUP), $0.990$ (CEG), $0.899$ (NS) and $0.977$ (Jazz), so $S^{PR}$ degenerates approximately into the degree product, which is precisely the PA baseline of this paper, the weakest of the seven local indices.

The conclusions of this subsection concern the \emph{unnormalized} PageRank scores of the original Charikhi protocol and the completion coefficient it prescribes; the full scope statement, and the reason this is not a general assessment of the PageRank centrality itself, are given in Section~6.1. Section~\ref{sec:ctrl} makes the distinction concrete through a set of control experiments on scale treatment: normalization is a necessary but not a sufficient condition, and non-redundancy with the local information is what actually decides whether the modulation mechanism takes effect.

\subsection{Reassessment results}

Table~\ref{tab:pr} reports the five-fold cross-validation AUCs of the seven local indices and their PR-MD fused counterparts on the eight datasets.

\begin{table}[!t]
	\centering
	\caption{AUC comparison between local similarity indices and PR-MD fused indices (fair protocol). The local-baseline rows and the PR-MD rows of this table are those of Table~\ref{tab:main}, repeated here so that the reassessment of Section~4 can be read without turning to it; the conclusions drawn from them are in Section~4.3. JC and SO are related by a monotone transformation of the common-neighbor count, hence their unfused AUCs coincide; both rows are reported for completeness (see Section~3.4 for their treatment in the BH family). Gains, means, win counts and gain intervals are computed from unrounded fold-level values; all entries are four-decimal displays, so a difference of two displayed cells need not equal the displayed gain}
	\footnotesize
	\setlength{\tabcolsep}{4pt}
	\begin{tabular}{@{}lcccccccc@{}}
		\toprule
		 & \textbf{BUP} & \textbf{CEG} & \textbf{NS} & \textbf{Jazz} & \textbf{Email} & \textbf{FB} & \textbf{GrQc} & \textbf{Wiki} \\
		\midrule
		CN & 0.8709 & 0.8288 & 0.9045 & 0.9503 & 0.9349 & 0.9905 & 0.9072 & 0.9193 \\
		PR-CN & 0.8611 & 0.8470 & 0.8928 & 0.9464 & 0.9374 & 0.9902 & 0.9072 & 0.9553 \\
		\midrule
		AA & 0.8808 & 0.8470 & 0.9072 & 0.9578 & 0.9401 & 0.9917 & 0.9076 & 0.9202 \\
		PR-AA & 0.8704 & 0.8566 & 0.8958 & 0.9570 & 0.9434 & 0.9915 & 0.9077 & 0.9555 \\
		\midrule
		RA & 0.8823 & 0.8505 & 0.9073 & 0.9664 & 0.9446 & 0.9929 & 0.9076 & 0.9200 \\
		PR-RA & 0.8719 & 0.8600 & 0.8959 & 0.9656 & 0.9478 & 0.9927 & 0.9078 & 0.9552 \\
		\midrule
		JC & 0.8571 & 0.7740 & 0.9027 & 0.9559 & 0.9239 & 0.9886 & 0.9072 & 0.9093 \\
		PR-JC & 0.8474 & 0.7841 & 0.8915 & 0.9551 & 0.9272 & 0.9884 & 0.9074 & 0.9446 \\
		\midrule
		HPI & 0.8681 & 0.7925 & 0.9041 & 0.9413 & 0.9036 & 0.9852 & 0.9069 & 0.8979 \\
		PR-HPI & 0.8602 & 0.8043 & 0.8936 & 0.9410 & 0.9074 & 0.9851 & 0.9071 & 0.9333 \\
		\midrule
		SO & 0.8571 & 0.7740 & 0.9027 & 0.9559 & 0.9239 & 0.9886 & 0.9072 & 0.9093 \\
		PR-SO & 0.8474 & 0.7841 & 0.8915 & 0.9551 & 0.9272 & 0.9884 & 0.9074 & 0.9446 \\
		\midrule
		PA & 0.6706 & 0.7599 & 0.6267 & 0.7656 & 0.8604 & 0.8311 & 0.7415 & 0.9332 \\
		PR-PA & 0.6710 & 0.7600 & 0.6284 & 0.7656 & 0.8604 & 0.8312 & 0.7426 & 0.9333 \\
		\bottomrule
	\end{tabular}
	\label{tab:pr}
\end{table}

As Table~\ref{tab:pr} shows, the change of PR-MD relative to the local baselines exhibits pronounced dataset dependence: small improvements on CEG and Email (e.g., CN rises from 0.8288 to 0.8470 on CEG, $+0.018$), but declines or stagnation on BUP, NS and Jazz (e.g., CN drops from 0.8709 to 0.8611 on BUP); Table~\ref{tab:pr} gives the full set of $7\times8$ index-by-network gains. Fold-level Wilcoxon signed-rank tests (40 pairs, one-sided; Table~\ref{tab:prtests}) show that the fusion differences of CN, AA, RA, JC and SO are all \textbf{non-significant} ($p$ between $0.12$ and $0.23$), while HPI approaches but does not reach significance ($p=0.061$); only PR-PA against PA shows a significant improvement of very small magnitude (mean difference $+0.0004$, $p=2.5\times10^{-5}$, BH-corrected $q=7.2\times10^{-5}$, 34/40 pairs positive). The dataset-level tests (8 networks) agree: except for PR-PA (Wilcoxon $p=0.020$, sign test $p=0.035$, 7/8 wins), the remaining six indices are all non-significant (Wilcoxon $p\geq 0.37$). We also re-examined PR fusion on the four original datasets over the full grid $\omega\in\{0,0.5,1,1.8,3,5,8,12\}$, $c\in\{0,0.1,0.3,1,3,5,10\}$, and found no unified parameter combination that significantly improves CN/AA/RA; the optimal value of the completion coefficient $c$ is always $0$ on these four networks, in all seven indices without exception (increasing $c$ only lifts the scores of low-degree negative pairs above those of the probe edges), whereas the original Charikhi setting takes $c=5$, a value that turns the mean $\Delta$AUC of the six common-neighbor indices negative, into the range $-0.0017\sim-0.0033$ (the first row of Table~\ref{tab:ctrl}). This conclusion is limited to the four medium-density networks above, in which the fraction of candidate pairs without common neighbors lies between $0.43$ and $0.97$: the optimal dose of the completion mechanism is set by the network structure, and on the near-tree-like wiki-Vote network the completion mechanism is precisely the only source of gain, as the next paragraph and Section~6.1 show.

\begin{table}[!t]
	\centering
	\caption{Statistical reassessment of the PR-fusion gains (PR-MD minus the local baseline) on the eight networks. \emph{Fold level}: one-sided Wilcoxon signed-rank test over the 40 fold pairs (8 networks $\times$ 5 folds, not independent). \emph{Dataset level}: one-sided Wilcoxon signed-rank test and one-sided sign test over the 8 network means (independent), the sidedness being prespecified from the directionality rationale of Section~3.4. \emph{Wins}: number of networks (out of 8) with a positive mean gain, counted on the unrounded fold-level means, so that a pair shown equal at four decimals may still be counted as a win. Only PA reaches significance at both levels (boldface).}
	\begin{tabular}{lccccc}
		\toprule
		Index & mean $\Delta$AUC & fold-level $p$ & dataset-level $p$ & sign $p$ & wins \\
		\midrule
		CN  & $+0.0039$ & $0.234$ & $0.473$ & $0.637$ & $4/8$ \\
		AA  & $+0.0032$ & $0.184$ & $0.527$ & $0.637$ & $4/8$ \\
		RA  & $+0.0032$ & $0.184$ & $0.527$ & $0.637$ & $4/8$ \\
		JC  & $+0.0034$ & $0.124$ & $0.473$ & $0.637$ & $4/8$ \\
		HPI & $+0.0041$ & $0.061$ & $0.371$ & $0.637$ & $4/8$ \\
		SO  & $+0.0034$ & $0.124$ & $0.473$ & $0.637$ & $4/8$ \\
		PA  & $+0.0004$ & $\mathbf{2.5\times10^{-5}}$ & $\mathbf{0.020}$ & $\mathbf{0.035}$ & $\mathbf{7/8}$ \\
		\bottomrule
	\end{tabular}
	\label{tab:prtests}
\end{table}

\textbf{Attribution of the wiki-Vote exception.} Table~\ref{tab:pr} shows that under the main protocol wiki-Vote gives a consistent gain of about $+0.036$ of PR-MD over the local baselines, the largest single-dataset gain of the paper. The mechanism behind this gain can be read off the data themselves: the gains of the six common-neighbor indices (CN/AA/RA/JC/HPI/SO) are highly consistent, all falling inside the interval $+0.0353\sim+0.0360$, whose width is $7\times10^{-4}$; whereas PA, which by the fusion rule~\eqref{eq:fusion} follows only the modulation branch and has no completion branch, gains merely $+5\times10^{-6}$ (the $+0.0001$ shown at four decimals in Table~\ref{tab:pr} is a rounding of that value). The lower end is contributed by AA and RA jointly: their gains agree to eight decimal places ($+0.0352954$), which is itself a stronger form of the same evidence, namely that the gain does not depend on the form of the local index at all. Because the gains are computed from the unrounded fold-level values, the interval need not coincide exactly with the differences of the displayed cells of Table~\ref{tab:pr}. This independence of the gain from the form of the local index is a direct corollary of the fact that the completion term $c\cdot S^{PR}$ does not depend on the form of the local index, whereas the multiplicative modulation term necessarily does; and the zero gain on PA amounts to a control experiment, carried out on real data, in which the completion mechanism is switched off. Together the two form an internal control test of the judgement that this gain is driven entirely by the completion mechanism.

The above results agree with the scale analysis of Section~4.2: \textbf{under the fair protocol, unnormalized PageRank fusion brings no consistent cross-network improvement over the local indices, and its single exception (wiki-Vote) is driven entirely by the completion mechanism}. This does not negate the conclusions reported in~\cite{A3} under their specific protocol and dataset composition, but shows that those conclusions are not robust across protocols, and that the practice of ``tuning $\omega_{MD}$ index by index'' lacks statistical justification. The comparison is also restricted in its network panel: of the eight networks of Table~\ref{tab:data}, only four are shared with the eleven-network panel of~\cite{A3}, so the present result bounds the reported improvement on the networks studied here rather than on that panel as a whole. This negative result is what motivates us to turn to DomiRank and to study parameter design principles systematically: an effective modulation mechanism requires the centrality product to have at once a dynamic range comparable to that of the local scores and an independent information content that does not overlap with the local information. Section~\ref{sec:ctrl} uses control experiments to show that the former is a necessary condition while the latter is the actual bottleneck. The wiki-Vote exception, in turn, points to another equally important design dimension: the gain of the completion mechanism is jointly determined by the network structure (the zero-score fraction) and the standalone predictive power of the centrality product; Section~6.1 develops the full ``dual-mechanism division of labor'' analysis accordingly.

\subsection{Control experiments on scale treatment: normalization is necessary but not sufficient}
\label{sec:ctrl}

The reassessment of Sections~4.1--4.3 changed two factors at once: the kind of centrality (PR versus DR) and the way its scale was handled (unnormalized versus min-max normalized). To separate the two we run a set of control experiments under an identical protocol (the same fold partitions, exhaustive candidate pairs, the same AUC rank-sum convention): all arms share the same fold partitions and differ only in the centrality vector, or in the way its scale is handled, so that the difference between arms can be attributed to that single variable. The experiment is carried out on the four original datasets covered by the PR full grid (BUP/CEG/NS/Jazz), giving $4\times5=20$ fold-level pairs. As a protocol-fidelity check, the original arm (Charikhi-type PR fusion with per-index $\omega_{MD}$ and $c=5$) and the DR arm ($\omega=5$, $c=0.1$) reproduce the corresponding values of Tables~\ref{tab:pr} and~\ref{tab:main} network by network, with a maximum deviation not exceeding $5\times10^{-5}$ in the cells that these control arms share with the main tables (the degree-corrected arm of Section~5.7 instead verifies the full table over all methods and networks, where the maximum absolute deviation is $6.8\times10^{-7}$), which guarantees that the new arms are comparable with the main-text numbers from the same source. The five arms are defined as follows.

(1) \textbf{PR-original arm}: $S^{PR}(x,y)=PR(x)\cdot PR(y)$ enters the fusion without any normalization, with per-index $\omega_{MD}$ and $c=5$, i.e., Eq.~\eqref{eq:prfusion};
(2) \textbf{PR order-preserving}: PR is first divided by its own maximum ($PR/\max PR$, an element-wise order-preserving transformation) and the product is then formed, with $\omega=5$ and $c=0.1$;
(3) \textbf{PR min-max}: PR is min-max normalized as in Eq.~\eqref{eq:minmax} and the product is then formed, with $\omega=5$ and $c=0.1$;
(4) \textbf{Normalized degree product}: the centrality product is replaced by the product of min-max normalized node degrees (that is, degree replaces centrality), with $\omega=5$ and $c=0.1$;
(5) \textbf{DR arm}: $S^{DR}$ is the min-max normalized DomiRank product, with $\omega=5$ and $c=0.1$.

\begin{table}[!t]
	\centering
	\caption{Control experiment on scale treatment: mean $\Delta$AUC of each arm over its own local baseline ($4$ original datasets $\times$ $5$ folds $=20$ fold-level pairs; bold marks a one-sided Wilcoxon signed-rank test with $p<0.05$)}
	\footnotesize
	\setlength{\tabcolsep}{2pt}
	\begin{tabular}{@{}lcccccccc@{}}
		\toprule
		Arm & CN & AA & RA & JC & HPI & SO & PA & Mean \\
		\midrule
		PR-original ($c=5$) & $-0.0018$ & $-0.0033$ & $-0.0033$ & $-0.0029$ & $-0.0017$ & $-0.0029$ & $\bm{+0.0005}$ & $-0.0022$ \\
		PR order-preserving ($C/\max C$) & $-0.0036$ & $-0.0063$ & $-0.0055$ & $+0.0017$ & $+0.0043$ & $+0.0019$ & $\bm{+0.0006}$ & $-0.0010$ \\
		PR min-max & $-0.0082$ & $-0.0107$ & $-0.0098$ & $-0.0030$ & $+0.0004$ & $-0.0028$ & $-0.0005$ & $-0.0050$ \\
		Normalized degree product & $-0.0126$ & $-0.0145$ & $-0.0134$ & $-0.0062$ & $-0.0028$ & $-0.0060$ & $-0.0004$ & $-0.0080$ \\
		DR arm ($c=0.1$) & $\bm{+0.0124}$ & $\bm{+0.0070}$ & $\bm{+0.0066}$ & $\bm{+0.0145}$ & $\bm{+0.0132}$ & $\bm{+0.0150}$ & $\bm{+0.0018}$ & $+0.0101$ \\
		\bottomrule
	\end{tabular}
	\label{tab:ctrl}
\end{table}

Table~\ref{tab:ctrl} supports three conclusions.

\textbf{First, scale calibration is a necessary condition.} The median magnitude of the unnormalized PR product over candidate pairs is only $6\times10^{-6}\sim8\times10^{-5}$, so the multiplicative factor $1+\omega S^{PR}$ is indistinguishable from $1$ and the modulation mechanism does not take effect at all (Section~4.2 already showed that with $c=0$ the value of $\omega$ exerts no influence on the result). After scale calibration (arms 2 and 3), the $\Delta$AUC of the three normalized indices JC/HPI/SO turns from negative to positive ($+0.0017$, $+0.0043$ and $+0.0019$ for the order-preserving arm), showing that the $\omega$ knob has indeed been activated.

\textbf{Second, scale calibration is far from sufficient.} The order-preserving arm obtains only $1/3$ to $1/9$ of the gain of the DR arm on JC/HPI/SO, and not one of the six common-neighbor indices reaches the $0.05$ level over the 20 pairs (the strongest is HPI with $p=0.088$); only PA is marginally significant ($p=0.045$), while the min-max and degree-product arms leave all seven indices non-significant ($p\geq0.26$). The normalized degree-product arm, obtained by replacing centrality entirely by node degree, closely resembles the PR min-max arm: when the centrality product is used as a predictor on its own, the two nearly coincide in full-candidate-set AUC ($0.6596$ versus $0.6553$ on BUP, $0.7599$ versus $0.7612$ on CEG, $0.6268$ versus $0.6256$ on NS, $0.7658$ versus $0.7551$ on Jazz, the degree product listed first), showing that the information carried by the scale-calibrated PR product overlaps strongly with degree information. The diagnostics explain this: on the four networks the Spearman rank correlation between node PageRank and degree reaches $0.899\sim0.990$, whereas for DR it is only $0.341\sim0.621$. When the centrality product is used as a predictor on its own, the AUC of the PR product ($0.64\sim0.77$) is in fact higher than that of the DR product ($0.64\sim0.69$); the PR product, however, is highly collinear with the PA baseline of this paper, whereas the DR product carries independent, non-degree information. What decides the success of fusion is therefore not the predictive accuracy of the centrality product itself, but whether an information gap that can complement the local score exists between the two.

\textbf{Third, the normalization scheme is itself a variable.} The order-preserving arm is uniformly better than the min-max arm on all seven indices (HPI, $+0.0043$ versus $+0.0004$; CN, $-0.0036$ versus $-0.0082$). The reason is that min-max is an affine map with a shift and does not preserve the ordering of the node product $C(x)C(y)$: on BUP, over the same candidate pairs, the AUC of the PR product used as a predictor falls from $0.6759$ for the original PR to $0.6553$ after min-max, whereas the order-preserving value is still $0.6759$, identical to the original PR, because $C/\max C$ is an element-wise order-preserving transformation that leaves the order of the node product unchanged. If the design intent is merely to calibrate the centrality product to a scale comparable with that of the local scores, therefore, an order-preserving scaling such as $C/\max C$ should be used; min-max introduces both a scale effect and an ordering effect, and the ordering effect is not necessarily what the designer intends.

The reassessment of Section~4 therefore reads: \textbf{the failure of unnormalized PageRank fusion is partly a matter of scale; but even after scale calibration, the high redundancy of the PageRank product with the local information still prevents it from delivering a consistent cross-network positive gain.} A centrality index can therefore serve as an effective fusion carrier only if it satisfies two conditions at the same time: (i) the scale of its product is comparable to that of the local scores, and (ii) its product is not strongly collinear with the local indices. Section~6.1 turns this into an operational two-step criterion.

\section{DomiRank-based fusion and parameter design}

\subsection{Method definition}

For unconnected nodes $v_x$, $v_y$, denote their DomiRank values (min-max normalized to $[0,1]$) by $\mathrm{DR}(x)$ and $\mathrm{DR}(y)$, and define the DomiRank similarity
\begin{equation}
S^{DR}(x,y) = \mathrm{DR}(x) \cdot \mathrm{DR}(y).
\end{equation}
Taking $C=\mathrm{DR}$ in Eq.~\eqref{eq:fusion} gives the DR-MD fused indices:
\begin{equation}
S_{DR}^{MD}(x,y) =
\begin{cases}
S^{MD}(x,y)\cdot \bigl(1 + \omega \cdot S^{DR}(x,y)\bigr), & |\Gamma_x \cap \Gamma_y| \neq 0 \text{ or } MD = PA, \\[6pt]
c \cdot S^{DR}(x,y), & |\Gamma_x \cap \Gamma_y| = 0 \text{ and } MD \neq PA,
\end{cases}
\label{eq:drfusion}
\end{equation}
where $(\omega,c)$ take unified values for all seven local indices. Unlike the PageRank product, the normalized $S^{DR}$ has a dynamic range of $O(10^{-2}\sim 1)$, so the multiplicative factor $1+\omega S^{DR}$ already produces substantive modulation of several to tens of percent at $\omega=O(1)$, and the dose of the completion term $c\cdot S^{DR}$ can be precisely controlled through $c$. DomiRank is computed with dominance threshold $\theta=1$, decay rate $\beta=1$ and relative competition intensity $\sigma=0.85$ (i.e., $0.85\sigma^{*}$; see Section~5.5 for the selection rationale).

\begin{figure}[!t]
	\centering
	\includegraphics[width=0.95\linewidth]{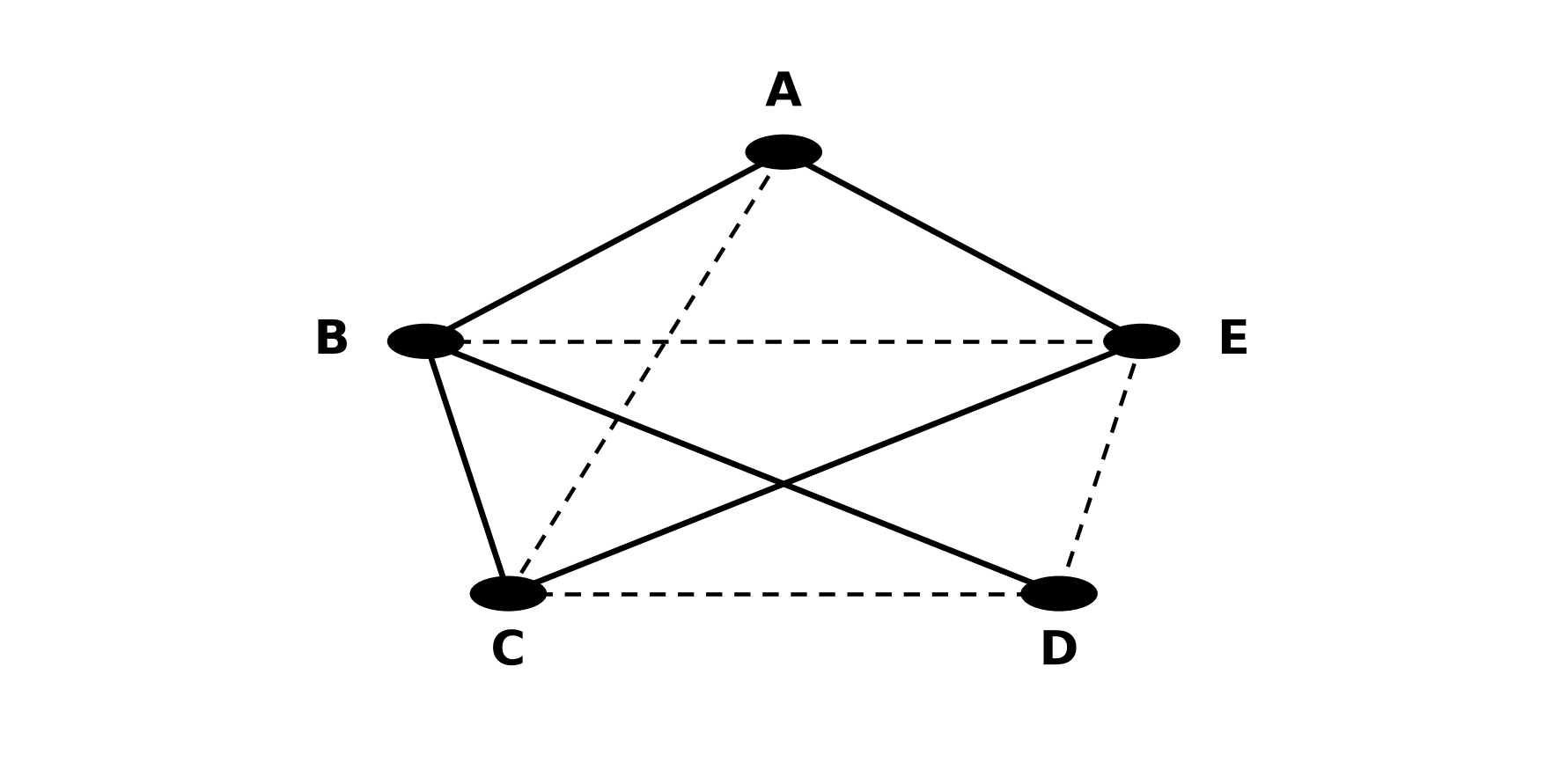}
	\caption{Schematic of the limitation of local similarity methods in link prediction}
	\label{fig:defect}
\end{figure}

\begin{figure}[!t]
	\centering
	\includegraphics[width=0.98\textwidth]{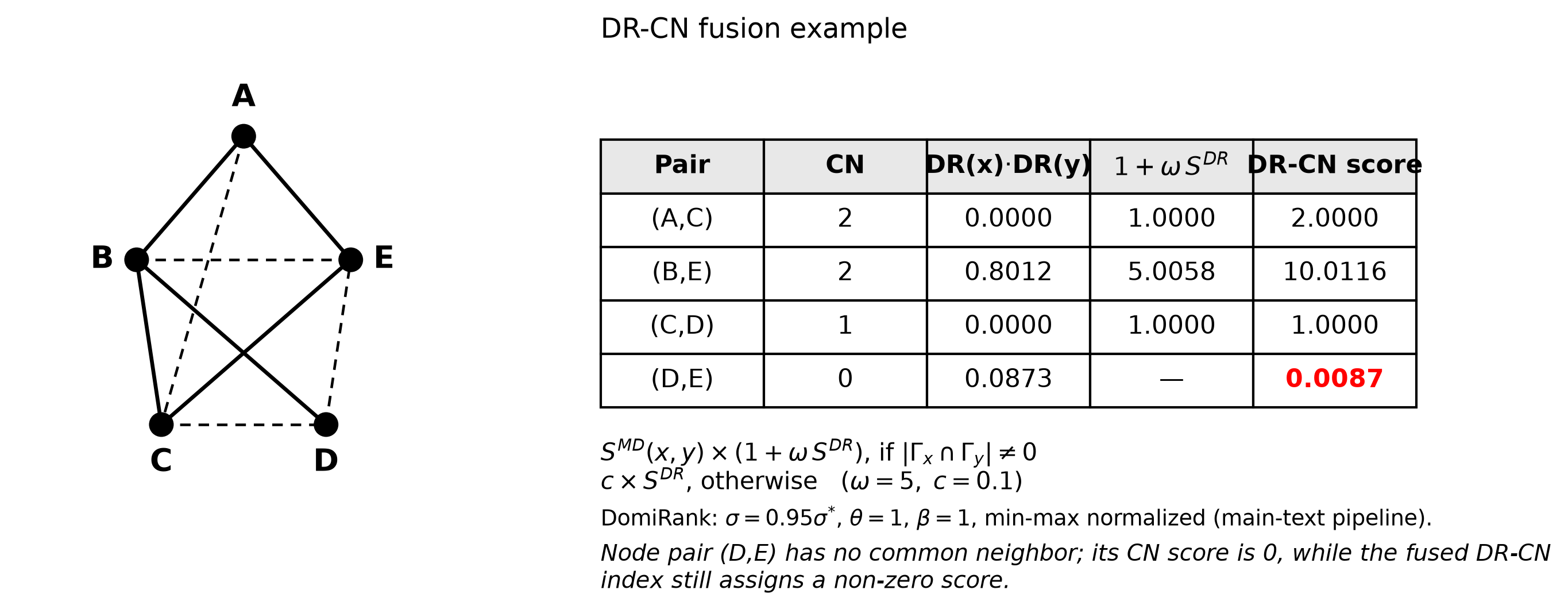}
	\caption{An illustrative example of the DR-CN fused index}
	\label{fig:example}
\end{figure}

The defect of local methods and the intuitive mechanism of fusion completion are illustrated in Figures~\ref{fig:defect} and~\ref{fig:example}: neither the node pair $(A,C)$ nor $(B,E)$ is directly connected, and different local methods assign significantly different or even uniformly zero scores; the DR-CN fused index, by contrast, still assigns a nonzero score to the pair $(D,E)$ without common neighbors, preserving predictive power. In this example DomiRank is computed with $\sigma=0.95\sigma^{*}$, $\theta=1$, $\beta=1$ and min-max normalization, following the main-text pipeline; $\sigma$ is taken at the upper end of the near-critical plateau (Section~5.5) because on this five-node graph a lower $\sigma$ would normalize the leaf node to zero and degenerate the completion example.

\subsection{Unified parameters: grid search and design space}

\begin{figure}[!t]
	\centering
	\includegraphics[width=0.98\textwidth]{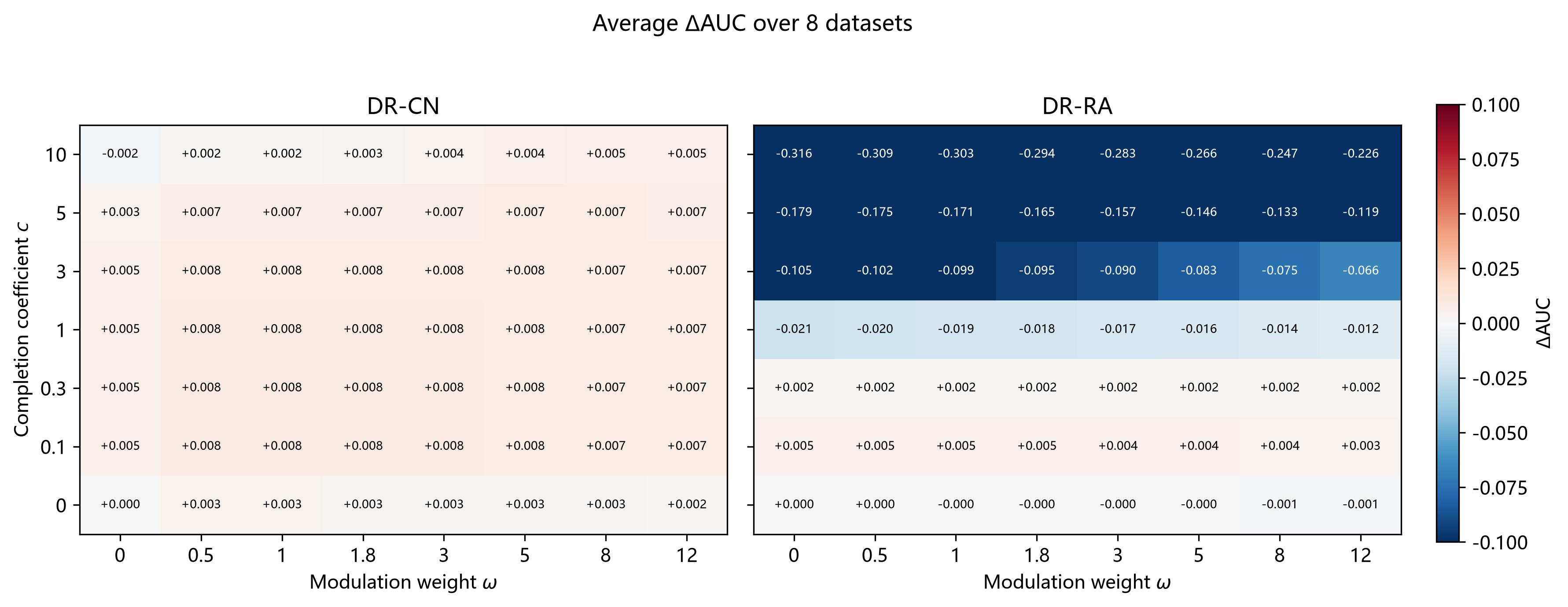}
	\caption{Average $\Delta$AUC of DR-CN and DR-RA over their local baselines on the $(\omega,c)$ plane (each cell averages over the 5 folds of the main protocol and the 8 networks, i.e., 40 fold-level values)}
	\label{fig:plane}
\end{figure}

To determine the unified parameters $(\omega,c)$ and characterize their design space, we evaluate the AUCs of the seven DR-MD indices over the 56 combinations of the following parameter grid. The two indices DR-CN and DR-RA are computed on all 40 folds of the eight datasets, and this is what Figure~\ref{fig:plane} plots; all seven DR-MD indices are computed on all 20 folds of the four original datasets covered by the PR full grid (BUP/CEG/NS/Jazz), and it is those that are used to select the unified parameters. The grid is
\begin{equation}
\omega\in\{0,\,0.5,\,1,\,1.8,\,3,\,5,\,8,\,12\},\qquad c\in\{0,\,0.1,\,0.3,\,1,\,3,\,5,\,10\}.
\end{equation}
Figure~\ref{fig:plane} shows the distribution of the mean $\Delta$AUC of DR-CN and DR-RA over their local baselines on the $(\omega,c)$ plane, which reveals three regularities:

(1) \textbf{The completion coefficient has a safety boundary}. The dose of the completion term must be strictly limited: for small-valued local indices such as RA, $\Delta$AUC is negative at every $c\geq1$ on the grid, for instance $-0.2663$ at $c=10$ for the operating point $\omega=5$ (the extreme over the $\omega$ grid being $-0.3155$ at $c=10$, $\omega=0$); for integer-valued indices such as CN, $c\leq 5$ still maintains a small positive gain, while $c=10$ turns slightly negative at $\omega=0$ ($-0.0024$). The mechanism is that an excessive centrality completion ranks a large number of negative pairs ``without common neighbors but with high centrality'' ahead of true probe edges, directly damaging AUC. Within the safe region ($0<c\leq 0.3$), the $\Delta$AUCs of DR-RA and DR-CN are both positive, and the mechanism decomposition (Section~5.6) shows that the remaining indices are likewise all positive at $c=0.1$.

(2) \textbf{The modulation weight exhibits a wide plateau, with a slight monotone slope inside it}. With $c=0.1$ fixed and $\omega$ increasing from 0.5 to 12, the two 8-network panel indices stay inside narrow bands, DR-CN within $+0.007\sim +0.008$ and DR-RA within $+0.003\sim +0.005$; and on the four original datasets used for parameter selection, the smallest per-index gain varies by no more than $0.0011$ over the whole interval (from $+0.00077$ to a peak of $+0.00187$), with no range in which performance deteriorates sharply. (The two coordinate systems are kept apart throughout: the bands above are panel means over the eight datasets and all 40 folds, whereas the per-index quantities, and the JC/SO/HPI/AA/RA values reported next, are means over the four original datasets and all 20 folds, since only those seven indices are computed on that grid. The two should not be combined into a single coordinate.) The fusion performance does not depend on the precise value of $\omega$, which provides direct evidence for the ``unified parameters'' claim and fundamentally avoids the overfitting risk of per-index tuning. The plateau is not strictly free of slope, however: it retains a slight monotone trend, with the gains of the normalized indices JC/SO/HPI rising monotonically with $\omega$ (JC from $+0.0089$ to $+0.0180$) and those of the count-based indices AA/RA falling monotonically (AA from $+0.0076$ to $+0.0058$, RA from $+0.0076$ to $+0.0052$; four-dataset values). This is self-consistent with the mechanism decomposition of Section~5.6, according to which the optimal $\omega$ for AA/RA lies near the left end of the plateau; the accurate statement is therefore that $\omega$ lies on a wide plateau rather than in a strictly peak-free region, but the slope inside the plateau is far smaller than the differences in gain between indices.

(3) \textbf{A warning about zero-completion degeneracy}. At $c=0$ (pure modulation), the $\Delta$AUC of DR-RA turns slightly negative ($-0.00047$ at $\omega=5$, DR fusion over all 40 folds of the eight datasets), showing that multiplicative modulation alone cannot sustain the improvement of all indices; small-dose completion is a necessary component of the fusion gain (see the mechanism decomposition in Section~5.6). This is the DR counterpart of the dead-knob finding of Section~4.2, and the two are complementary rather than conflicting: Section~4.2 shows that the \emph{unnormalized PR product} leaves the fused score unchanged across a $24\times$ range of $\omega$ (seven-index mean $+0.00068$ at every $\omega$, over the four original datasets), whereas here the normalized DR product does respond to $\omega$ (seven-index mean rising from $+0.00174$ to $+0.00430$ over the same range, four original datasets) but the response is not by itself sufficient to keep every index above its baseline, because the pure modulation branch leaves the zero-score node pairs unordered.

The unified parameters are determined by the \textbf{maximin} criterion, i.e., by maximizing the smallest per-index gain. The commonly used ``maximal average gain'' criterion does not give a robust location on our grid: ranking the 56 combinations by the seven-index mean $\Delta$AUC over the four original datasets singles out $(\omega,c)=(12,0.1)$ (mean $+0.01068$), but that point lies on the upper boundary of the $\omega$ grid and its neighbourhood has already begun to fall off ($+0.01049$ at $\omega=8$), whereas under the maximin criterion the optimum is $(\omega,c)=(8,0.1)$ (smallest per-index gain $+0.00187$; the criterion is well posed over the entire grid without any restriction to a safe region, and it is attained at two tied points, $(8,0.1)$ and $(8,0.3)$, the neighbouring points $(5,0.1)$ and $(5,0.3)$ following at $+0.00185$). Maximin is the more appropriate criterion because the two families of indices prefer opposite directions of $\omega$ (regularity (2)): the average maximum would be dominated by the large gains of the normalized indices, whereas maximin captures this trade-off explicitly. We take
\begin{equation}
\omega = 5, \qquad c = 0.1.
\label{eq:params}
\end{equation}
Its smallest per-index gain is $+0.00185$, only about $2.6\times10^{-5}$ below the maximin optimum; it ranks $3/56$ under this criterion while lying inside the grid, which avoids the boundary extrapolation at $\omega=12$. The price of choosing $\omega=5$ is quantifiable: relative to $\omega=8$ it slightly sacrifices the optimality of AA/RA, the net contribution of the modulation mechanism to the two being of order $-0.0001$ and $-0.0004$ (Table~\ref{tab:decomp}), far below the fold-level fluctuation (the fold-level standard deviation of these two indices is about $0.002$, and one-sided Wilcoxon tests give $p=0.31$ and $0.90$, indistinguishable from $0$). What is gained is the substantive methodological benefit of not tuning per index, while the completion mechanism ($c=0.1$) supplies a universal gain base of about $+0.0046\sim +0.0049$ for the six common-neighbor indices (Section~5.6); together the two still give all seven indices a significant positive gain at $\omega=5$.
These parameters are selected on the same fold partitions used for performance evaluation, so the reported values carry a mild selection bias. Three pieces of evidence indicate that this bias does not affect the validity of the conclusions: first, as regularity (2) states, the conclusions hold over the entire plateau $\omega\in[0.5,12]\times c\in[0.1,0.3]$ and do not depend on the precise values of Eq.~\eqref{eq:params}; second, the mechanism decomposition of Section~5.6 shows that even with $\omega=0$ (the modulation mechanism switched off entirely), the completion mechanism still produces significant gains; third, we further conduct \textbf{leave-one-network-out validation} to directly test the transferability of the unified parameters: each time one dataset is held out, parameters are selected by mean $\Delta$AUC on the fold-level grid $\omega\times c\in\{1,3,5,8\}\times\{0.05,0.1,0.3\}$ of the remaining seven datasets, and then used to evaluate the held-out network; the results are reported at the end of Section~5.3. This validation differs from the main selection rule in three respects: it uses the mean-$\Delta$AUC criterion instead of maximin, a coarser grid (12 instead of 56 combinations, still covering the plateau interior $[1,8]\times[0.05,0.3]$ identified by regularity (2)), and only the representative index DR-RA rather than all seven. The reasons are computational and diagnostic: a maximin selection over all seven indices on each of the eight leave-one-out problems would require $8\times 7\times 12$ full seven-index evaluations, and the deployment question, namely whether a selection rule fitted on other networks transfers, is answered most economically for the family's representative index. We regard the criterion/grid mismatch as part of the selection-bias limitation (Section~6.2), where fully nested cross-validation is listed as the rigorous confirmation scheme.

\clearpage

\subsection{Main results}

\begin{figure}[!t]
	\centering
	\includegraphics[width=0.9\textwidth]{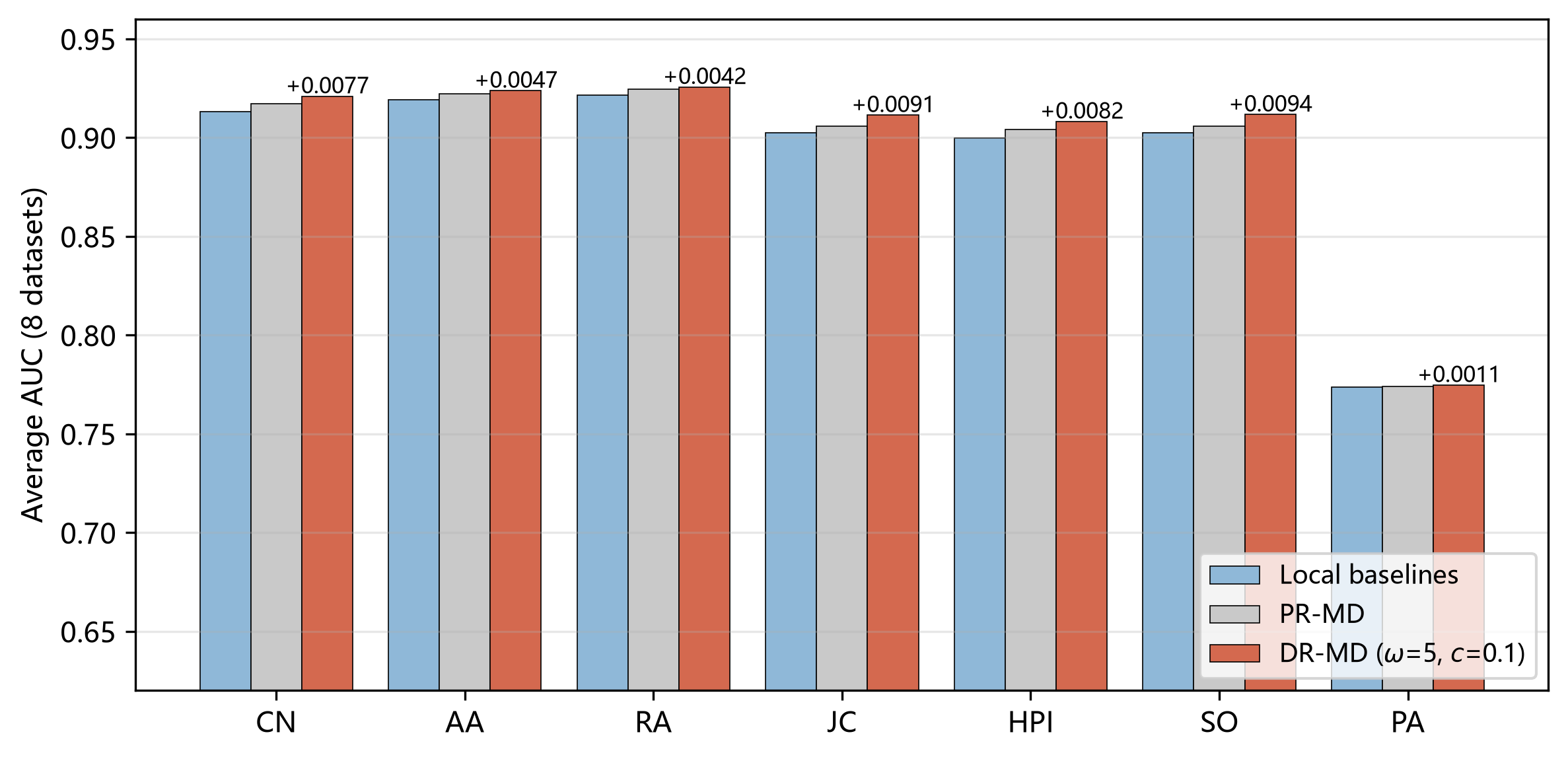}
	\caption{Average AUC of local baselines, PR-MD and DR-MD over the 8 datasets (each value averages over the 5 folds per dataset, and each bar further averages over the 8 datasets, i.e., 40 fold-level values; numbers on top of bars are the gains of DR-MD over the baselines)}
	\label{fig:avg}
\end{figure}

Under the unified parameters~\eqref{eq:params}, the five-fold cross-validation results of the seven DR-MD indices, the local baselines and the PR-MD indices are listed in Table~\ref{tab:main}; the significance tests are summarized in Table~\ref{tab:wilcox}; a visual comparison of the average AUCs is shown in Figure~\ref{fig:avg}.

\begin{table}[!t]
	\centering
	\caption{AUC of local baselines, PR-MD and DR-MD (unified parameters $\omega=5$, $c=0.1$). JC and SO are related by a monotone transformation of the common-neighbor count, hence their unfused AUCs coincide; both rows are reported for completeness (see Section~3.4 for their treatment in the BH family)}
	\footnotesize
	\setlength{\tabcolsep}{4pt}
	\begin{tabular}{@{}lcccccccc@{}}
		\toprule
		 & \textbf{BUP} & \textbf{CEG} & \textbf{NS} & \textbf{Jazz} & \textbf{Email} & \textbf{FB} & \textbf{GrQc} & \textbf{Wiki} \\
		\midrule
		CN & 0.8709 & 0.8288 & 0.9045 & 0.9503 & 0.9349 & 0.9905 & 0.9072 & 0.9193 \\
		PR-CN & 0.8611 & 0.8470 & 0.8928 & 0.9464 & 0.9374 & 0.9902 & 0.9072 & 0.9553 \\
		DR-CN & \textbf{0.8839} & \textbf{0.8585} & \textbf{0.9150} & 0.9465 & \textbf{0.9388} & \textbf{0.9916} & \textbf{0.9215} & 0.9117 \\
		\midrule
		AA & 0.8808 & 0.8470 & 0.9072 & 0.9578 & 0.9401 & 0.9917 & 0.9076 & 0.9202 \\
		PR-AA & 0.8704 & 0.8566 & 0.8958 & 0.9570 & 0.9434 & 0.9915 & 0.9077 & 0.9555 \\
		DR-AA & \textbf{0.8885} & \textbf{0.8627} & \textbf{0.9167} & 0.9530 & \textbf{0.9425} & \textbf{0.9925} & \textbf{0.9219} & 0.9122 \\
		\midrule
		RA & 0.8823 & 0.8505 & 0.9073 & 0.9664 & 0.9446 & 0.9929 & 0.9076 & 0.9200 \\
		PR-RA & 0.8719 & 0.8600 & 0.8959 & 0.9656 & 0.9478 & 0.9927 & 0.9078 & 0.9552 \\
		DR-RA & \textbf{0.8895} & \textbf{0.8654} & \textbf{0.9168} & 0.9614 & \textbf{0.9466} & \textbf{0.9938} & \textbf{0.9220} & 0.9099 \\
		\midrule
		JC & 0.8571 & 0.7740 & 0.9027 & 0.9559 & 0.9239 & 0.9886 & 0.9072 & 0.9093 \\
		PR-JC & 0.8474 & 0.7841 & 0.8915 & 0.9551 & 0.9272 & 0.9884 & 0.9074 & 0.9446 \\
		DR-JC & \textbf{0.8761} & \textbf{0.7992} & \textbf{0.9131} & \textbf{0.9593} & \textbf{0.9303} & \textbf{0.9896} & \textbf{0.9217} & 0.9021 \\
		\midrule
		HPI & 0.8681 & 0.7925 & 0.9041 & 0.9413 & 0.9036 & 0.9852 & 0.9069 & 0.8979 \\
		PR-HPI & 0.8602 & 0.8043 & 0.8936 & 0.9410 & 0.9074 & 0.9851 & 0.9071 & 0.9333 \\
		DR-HPI & \textbf{0.8857} & \textbf{0.8138} & \textbf{0.9148} & \textbf{0.9443} & \textbf{0.9085} & \textbf{0.9862} & \textbf{0.9214} & 0.8906 \\
		\midrule
		SO & 0.8571 & 0.7740 & 0.9027 & 0.9559 & 0.9239 & 0.9886 & 0.9072 & 0.9093 \\
		PR-SO & 0.8474 & 0.7841 & 0.8915 & 0.9551 & 0.9272 & 0.9884 & 0.9074 & 0.9446 \\
		DR-SO & \textbf{0.8773} & \textbf{0.8001} & \textbf{0.9133} & \textbf{0.9591} & \textbf{0.9305} & \textbf{0.9897} & \textbf{0.9217} & 0.9022 \\
		\midrule
		PA & 0.6706 & 0.7599 & 0.6267 & 0.7656 & 0.8604 & 0.8311 & 0.7415 & 0.9332 \\
		PR-PA & 0.6710 & 0.7600 & 0.6284 & 0.7656 & 0.8604 & 0.8312 & 0.7426 & 0.9333 \\
		DR-PA & \textbf{0.6739} & \textbf{0.7623} & \textbf{0.6298} & 0.7642 & \textbf{0.8605} & \textbf{0.8313} & \textbf{0.7425} & 0.9331 \\
		\bottomrule
	\end{tabular}
	\label{tab:main}
\end{table}

\begin{table}[!t]
	\centering
	\caption{Summary of fold-level Wilcoxon signed-rank tests (40 pairs, one-sided, alternative: the former is better; $q$ values are BH-FDR corrected over the 31-test BH family) and dataset-level tests}
	\footnotesize
	\setlength{\tabcolsep}{3pt}
	\begin{tabular}{@{}lccccccl@{}}
		\toprule
		Comparison & Mean $\Delta$AUC & Wins/40 & $p$-value & $q$-value & Wins/8 & Wilcoxon $p$ & Sign $p$ \\
		\midrule
		DR-CN vs CN & $+0.0077$ & 29/40 & $2.3\times10^{-4}$ & $5.4\times10^{-4}$ & 6/8 & $5.5\times10^{-2}$ & $1.4\times10^{-1}$ \\
		DR-AA vs AA & $+0.0047$ & 29/40 & $1.7\times10^{-3}$ & $2.9\times10^{-3}$ & 6/8 & $9.8\times10^{-2}$ & $1.4\times10^{-1}$ \\
		DR-RA vs RA & $+0.0042$ & 29/40 & $3.3\times10^{-3}$ & $4.9\times10^{-3}$ & 6/8 & $1.2\times10^{-1}$ & $1.4\times10^{-1}$ \\
		DR-JC vs JC & $+0.0091$ & 34/40 & $3.5\times10^{-6}$ & $1.3\times10^{-5}$ & 7/8 & $2.7\times10^{-2}$ & $3.5\times10^{-2}$ \\
		DR-HPI vs HPI & $+0.0082$ & 34/40 & $5.6\times10^{-6}$ & $1.9\times10^{-5}$ & 7/8 & $2.7\times10^{-2}$ & $3.5\times10^{-2}$ \\
		DR-SO vs SO & $+0.0094$ & 34/40 & $2.3\times10^{-6}$ & $1.0\times10^{-5}$ & 7/8 & $2.7\times10^{-2}$ & $3.5\times10^{-2}$ \\
		DR-PA vs PA & $+0.0011$ & 28/40 & $1.5\times10^{-3}$ & $2.8\times10^{-3}$ & 6/8 & $7.4\times10^{-2}$ & $1.4\times10^{-1}$ \\
		\midrule
		DR-CN vs PR-CN & $+0.0038$ & 32/40 & $1.9\times10^{-3}$ & $3.0\times10^{-3}$ & -- & -- & -- \\
		DR-AA vs PR-AA$^{\mathrm{n.s.}}$ & $+0.0016$ & 24/40 & $6.4\times10^{-2}$ & $8.6\times10^{-2}$ & -- & -- & -- \\
		DR-RA vs PR-RA$^{\mathrm{n.s.}}$ & $+0.0010$ & 24/40 & $9.6\times10^{-2}$ & $1.2\times10^{-1}$ & -- & -- & -- \\
		DR-JC vs PR-JC & $+0.0057$ & 35/40 & $9.8\times10^{-4}$ & $2.0\times10^{-3}$ & -- & -- & -- \\
		DR-HPI vs PR-HPI & $+0.0042$ & 35/40 & $1.3\times10^{-3}$ & $2.4\times10^{-3}$ & -- & -- & -- \\
		DR-SO vs PR-SO & $+0.0060$ & 35/40 & $9.8\times10^{-4}$ & $2.0\times10^{-3}$ & -- & -- & -- \\
		DR-PA vs PR-PA$^{\mathrm{n.s.}}$ & $+0.0007$ & 24/40 & $6.9\times10^{-2}$ & $9.0\times10^{-2}$ & -- & -- & -- \\
		\midrule
		DR-RA vs Katz & $+0.0088$ & 33/40 & $1.8\times10^{-3}$ & $3.0\times10^{-3}$ & -- & -- & -- \\
		DR-RA vs RWR & $-0.0003$ & 25/40 & $5.4\times10^{-1}$ & $5.4\times10^{-1}$ & -- & -- & -- \\
		DR-RA vs SimRank & $+0.0547$ & 40/40 & $9.1\times10^{-13}$ & $2.8\times10^{-11}$ & -- & -- & -- \\
		DR-RA vs LNB & $+0.0077$ & 34/40 & $9.8\times10^{-6}$ & $3.0\times10^{-5}$ & -- & -- & -- \\
		DR-RA vs CN2D & $+0.0165$ & 35/40 & $2.2\times10^{-8}$ & $1.4\times10^{-7}$ & -- & -- & -- \\
		DR-RA vs CNC & $+0.0240$ & 35/40 & $1.9\times10^{-7}$ & $1.0\times10^{-6}$ & -- & -- & -- \\
		DR-RA vs CND & $+0.0252$ & 35/40 & $7.9\times10^{-9}$ & $8.2\times10^{-8}$ & -- & -- & -- \\
		DR-RA vs CCPA & $+0.0130$ & 33/40 & $2.9\times10^{-5}$ & $7.6\times10^{-5}$ & -- & -- & -- \\
		DR-RA vs Gravity & $+0.0702$ & 35/40 & $1.2\times10^{-10}$ & $1.9\times10^{-9}$ & -- & -- & -- \\
		DR-RA vs CNPop & $+0.0152$ & 34/40 & $1.7\times10^{-8}$ & $1.3\times10^{-7}$ & -- & -- & -- \\
		\bottomrule
	\end{tabular}
	\label{tab:wilcox}
\end{table}

Three observations follow from Tables~\ref{tab:main} and~\ref{tab:wilcox}. (1) \textbf{All seven DR-MD indices achieve statistically significant fold-level improvements over their local baselines} ($p\leq 3.3\times10^{-3}$, BH-corrected $q\leq 4.9\times10^{-3}$), with mean gains ranging from $+0.0011$ for PA to $+0.0094$ for SO; 34 of the 40 pairs are strictly positive for JC/HPI/SO, and 28--29 for CN/AA/RA/PA. The dataset-level tests (8 networks) further show that JC/HPI/SO are significant under both the Wilcoxon and sign tests ($p=0.027/0.035$), while the gains of CN/AA/RA/PA do not reach significance at the dataset level (Wilcoxon $p$ between $0.055$ and $0.12$), indicating that the fold-level significance mainly stems from within-fold consistency and that consistent cross-network wins concentrate on the normalized indices; this reading agrees with the mechanism decomposition of Section~5.6. Accordingly, throughout this paper we take the dataset-level tests (independent network units) as the primary significance evidence and the fold-level tests as supporting evidence. (2) \textbf{DR-MD is significantly better than PR-MD on the four indices CN/JC/HPI/SO} (fold-level $p\leq 1.9\times10^{-3}$), while the differences on AA/RA/PA are non-significant ($p$ between $6.4\times10^{-2}$ and $9.6\times10^{-2}$, marked $^{\mathrm{n.s.}}$ in Table~\ref{tab:wilcox}); combined with the reassessment of Section~4, the essence of this advantage is that the dynamic range of the DomiRank product makes the fusion mechanism genuinely effective, whereas the PageRank product does not constitute effective modulation. (3) The magnitude of the gain varies with the type of local index: it is largest for the normalized indices (JC/SO/HPI), intermediate for the integer/weighted-count indices (CN/AA/RA), and weakest for PA. The explanation of observation (3) agrees with the mechanism decomposition of Section~5.6: PA itself already carries global information in the form of a ``degree product'', so the marginal increment of centrality completion is naturally minimal; the source of the fusion gain is the ability to distinguish node pairs ``without common neighbors but with differentiated node importance'', rather than an equal amplification of all indices. All of the above statements refer to the exhaustive-negative-sample protocol; Section~\ref{sec:degcorr} shows that the gains over the local baselines do not survive a degree-corrected benchmark, in which positive and negative pairs are matched in degree.

DR-MD does not exceed the baselines on every dataset: on Jazz for CN/AA/RA/PA, and on Wiki for all seven indices, DR-MD fails to exceed the local baselines (e.g., DR-CN on Jazz is 0.9465, below the CN baseline 0.9503; DR-RA on Wiki is 0.9099, below the RA baseline 0.9200). Jazz has an average degree as high as 27.70, so the vast majority of node pairs in the training graph possess common neighbors and the coverage of the completion term is small; Wiki is the largest network (7066 nodes) with a clustering coefficient of only 0.142 and a near-tree-like structure, where the correlation between node-centrality differences and local ordering is weak, so multiplicative modulation slightly disturbs the local ordering on both networks. This suggests that the applicable scenario of the fusion strategy is networks of medium or lower density and medium scale, consistent with the original motivation of ``alleviating the zero-score failure in sparse networks''. The cross-comparison on Wiki is treated separately (see Section~4.3): on this network, PR-MD is instead markedly better than DR-MD (e.g., PR-CN is 0.9553 while DR-CN is only 0.9117): the near-identity modulation of PR preserves the local ordering while its completion term provides a high-quality ordering, whereas the genuinely effective DR modulation disturbs the ordering. The same network, two centralities, and opposite net effects of the two mechanisms provide direct evidence for the ``dual-mechanism division of labor'' of Section~6.1.

\textbf{Leave-one-network-out validation} further tests the transferability of the unified parameters. The protocol is the one specified in Section~5.2 (parameters selected on seven networks, evaluated on the eighth; mean-$\Delta$AUC criterion on the grid $\omega\times c\in\{1,3,5,8\}\times\{0.05,0.1,0.3\}$; representative index DR-RA); the results are shown in Table~\ref{tab:loo}. Six of the eight networks still achieve positive gains under the leave-one-out protocol ($+0.0008\sim+0.0144$), with GrQc the largest ($+0.0144$); Jazz ($w^{*}=3$) and Wiki ($c^{*}=0.1$) show slight negative gains ($-0.0032$ and $-0.0103$), consistent in direction with their negative gains under the unified parameters. The leave-one-out validation shows that the selection bias of the unified parameters only mildly affects a few high-density/large-scale networks, and the overall conclusion (DR fusion is effective on most networks) is not dominated by the selection bias.

\begin{table}[!t]
	\centering
	\caption{Leave-one-network-out validation: parameters selected on the remaining 7 networks and evaluated on the held-out network (DR-RA)}
	\footnotesize
	\setlength{\tabcolsep}{6pt}
	\begin{tabular}{@{}lcccccc@{}}
		\toprule
		Held-out network & $\omega^{*}$ & $c^{*}$ & DR-RA AUC & RA baseline & $\Delta$AUC \\
		\midrule
		BUP & 1 & 0.05 & 0.8901 & 0.8823 & $+0.0078$ \\
		CEG & 1 & 0.05 & 0.8642 & 0.8505 & $+0.0137$ \\
		NS & 1 & 0.05 & 0.9168 & 0.9073 & $+0.0095$ \\
		Jazz & 3 & 0.05 & 0.9632 & 0.9664 & $-0.0032$ \\
		Email & 1 & 0.05 & 0.9467 & 0.9446 & $+0.0021$ \\
		FB & 1 & 0.05 & 0.9937 & 0.9929 & $+0.0008$ \\
		GrQc & 1 & 0.05 & 0.9220 & 0.9076 & $+0.0144$ \\
		Wiki & 1 & 0.1 & 0.9097 & 0.9200 & $-0.0103$ \\
		\bottomrule
	\end{tabular}
	\label{tab:loo}
\end{table}

\textbf{Repeated cross-validation.} Finally, to test the robustness of the above conclusions to fold-partition randomness, we repeat the complete five-fold cross-validation 10 times with independent random seeds for every network (50 partitions per network), evaluating the representative indices CN, RA, their DR fused counterparts, Katz and RWR. The results agree closely with the single-partition ones. (1) The mean gains of DR-CN and DR-RA over the local baselines across the 50 partitions have, on all eight networks, exactly the same signs as in Table~\ref{tab:main} (positive on BUP/CEG/NS/Email/FB/GrQc, negative on Jazz and Wiki), and the magnitudes match closely (e.g., DR-CN on CEG: $+0.0259$ versus $+0.0297$; DR-RA on Wiki: $-0.0097$ versus $-0.0101$); across the 10 repetitions, the repetition-mean gains are positive in all 10 repetitions on the six networks and in none on Jazz and Wiki. (2) Tests with the eight repetition means as paired units give DR-CN over CN (mean $+0.0076$, Wilcoxon $p=0.055$) and DR-RA over RA (mean $+0.0044$, Wilcoxon $p=0.098$), almost coinciding with the dataset-level conclusions of the single partition ($p=0.055$ and $0.12$, Table~\ref{tab:wilcox}); the two columns use the same unit in both cases, namely the eight network means, so that the repetition means differ from the single-partition means only through the fold partition and not through the choice of test unit; the fold-level tests pooling all 400 partitions give $p\leq1.2\times10^{-17}$, showing that the fold-level significance is not an artifact of a single partition. (3) The per-network win/loss relations of DR-RA against Katz and RWR agree with Section~5.4 (per network, RWR wins all 10 repetitions on BUP/CEG/Wiki and DR-RA wins all repetitions on the remaining networks; in the panel average over the eight networks the repetition mean of DR-RA against RWR is $-0.0022$, still comparable); the only change is that the small single-partition advantage of DR-RA over Katz on BUP ($+0.0018$) does not survive repetition (Katz wins all 10 repetitions, repetition mean $-0.0101$), which shows that single-partition conclusions on small networks are susceptible to partition randomness and supports the use of cross-network pooled tests as the main evidence in this section.

\subsection{Comparison with global and advanced methods}

\begin{figure}[!t]
	\centering
	\includegraphics[width=0.98\textwidth]{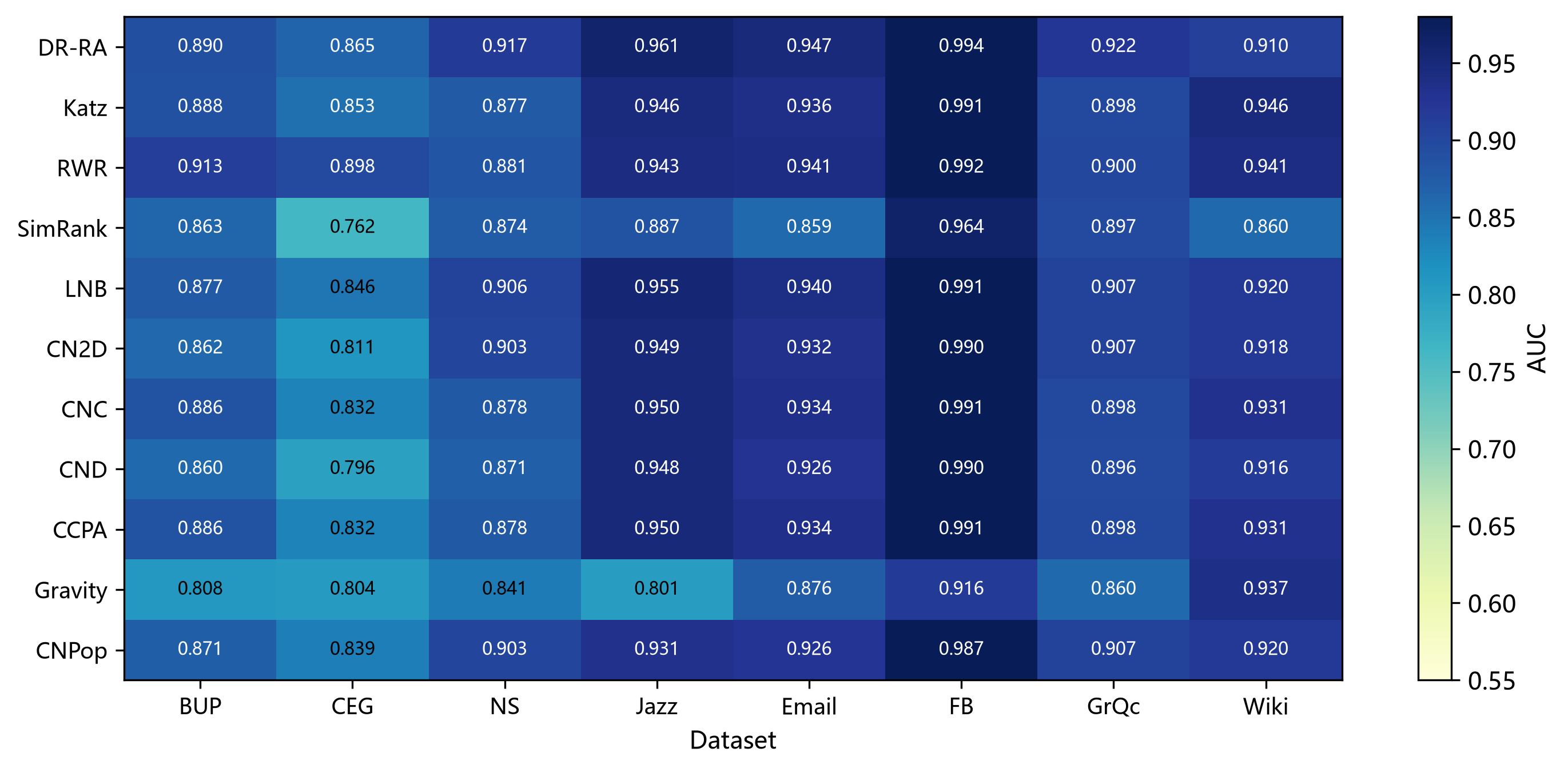}
	\caption{AUC of DR-RA versus global/advanced indices on the 8 datasets}
	\label{fig:global}
\end{figure}

We compare DR-RA, the most competitive member of the fusion family, with ten global and advanced indices. Parameter settings: Katz with $\alpha=0.001$; RWR with restart probability $c=0.85$; SimRank with damping factor $r=0.8$; CNC with $\alpha=0.8$ (Jaccard-normalized common neighbors); CN2D with $\alpha=0.004$; CCPA, Gravity and CNPop are implemented in the reported forms of Section~3.2. We deliberately keep all baselines at the parameter values reported or recommended in their original papers, rather than tuning them per fold on our evaluation partitions: grid-tuning every baseline on the same folds used for evaluation would couple the comparison to the partition and contradict the fair-protocol premise of this paper. The price of this choice is an asymmetry with our own grid-searched $(\omega,c)$, whose selection bias is analyzed and controlled in Section~5.2 (plateau property, leave-one-network-out validation) and Section~5.3 (repeated cross-validation). The asymmetry is one-sided and its direction is known: our $(\omega,c)$ is chosen to maximize the gain, whereas each baseline is evaluated at one fixed parameter value that may be anywhere on its own performance profile. The resulting bias can only favour DR-RA, so every comparison reported here is an upper bound on the advantage that a fully tuned baseline would face. The bound is not quantified in this paper, because that would require a per-fold parameter sweep of Katz, RWR, CN2D and CNC on the evaluation partitions; one calibration point is available from the repeated cross-validation of Section~5.3, where the single-partition advantage of DR-RA over Katz on BUP ($+0.0018$) does not survive repetition (Katz wins all ten repetitions, repetition mean $-0.0101$), indicating that the Katz baseline is not at a disadvantage from an unfavorable parameter choice on that network. All comparisons of this subsection should be read under this declared parameter-setting asymmetry. The results are shown in Table~\ref{tab:global} and Figure~\ref{fig:global}.

\begin{table}[!t]
	\centering
	\caption{AUC comparison between DR-RA (unified parameters) and global/advanced indices}
	\footnotesize
	\setlength{\tabcolsep}{4pt}
	\begin{tabular}{@{}lccccccccc@{}}
		\toprule
		 & \textbf{BUP} & \textbf{CEG} & \textbf{NS} & \textbf{Jazz} & \textbf{Email} & \textbf{FB} & \textbf{GrQc} & \textbf{Wiki} & Mean \\
		\midrule
		DR-RA & 0.8895 & 0.8654 & 0.9168 & 0.9614 & 0.9466 & 0.9938 & 0.9220 & 0.9099 & 0.9257 \\
		Katz & 0.8877 & 0.8530 & 0.8771 & 0.9458 & 0.9359 & 0.9913 & 0.8983 & 0.9462 & 0.9169 \\
		RWR & \textbf{0.9126} & \textbf{0.8975} & 0.8814 & 0.9430 & 0.9405 & 0.9915 & 0.9004 & 0.9410 & 0.9260 \\
		SimRank & 0.8634 & 0.7620 & 0.8745 & 0.8870 & 0.8591 & 0.9641 & 0.8974 & 0.8602 & 0.8710 \\
		LNB & 0.8776 & 0.8469 & 0.9063 & 0.9547 & 0.9396 & 0.9909 & 0.9072 & 0.9204 & 0.9180 \\
		CN2D & 0.8624 & 0.8111 & 0.9028 & 0.9492 & 0.9323 & 0.9903 & 0.9070 & 0.9184 & 0.9092 \\
		CNC & 0.8724 & 0.7776 & 0.8762 & 0.9555 & 0.9232 & 0.9895 & 0.8984 & 0.9207 & 0.9017 \\
		CND & 0.8603 & 0.7964 & 0.8707 & 0.9483 & 0.9264 & 0.9898 & 0.8960 & 0.9161 & 0.9005 \\
		CCPA & 0.8862 & 0.8324 & 0.8780 & 0.9499 & 0.9342 & 0.9915 & 0.8983 & 0.9307 & 0.9126 \\
		Gravity & 0.8081 & 0.8040 & 0.8414 & 0.8012 & 0.8759 & 0.9162 & 0.8604 & 0.9368 & 0.8555 \\
		CNPop & 0.8715 & 0.8386 & 0.9033 & 0.9314 & 0.9261 & 0.9869 & 0.9067 & 0.9195 & 0.9105 \\
		\bottomrule
	\end{tabular}
	\label{tab:global}
\end{table}

As Table~\ref{tab:global} and the third part of Table~\ref{tab:wilcox} show, DR-RA achieves statistically significant advantages over nine of the ten comparison methods, namely Katz, SimRank, LNB, CN2D, CNC, CND, CCPA, Gravity and CNPop, under the fixed parameter settings reported in the original literature (Section~5.4) ($p\leq 1.8\times10^{-3}$, BH-corrected $q\leq 3.0\times10^{-3}$), with average AUC higher by $0.0077\sim 0.0702$. The margin over SimRank is the widest of the nine ($+0.0547$ in panel average, DR-RA ahead in all $40$ fold-level pairs) and should be read in light of that method's scope: SimRank is defined on directed graphs through $In(x)$ and is placed at an inherent disadvantage on the undirected panels used here, so this gap measures the choice of comparison method at least as much as the merit of the fusion rule. Against RWR it is overall tied (mean $-0.0003$, $p=0.54$), and per dataset RWR is better on BUP, CEG and Wiki, while DR-RA is better on NS, Jazz, Email, FB and GrQc. CNC and CCPA are ordered differently here, for the reason given in Section~3.2: the Jaccard normalization in CNC breaks the order equivalence that a raw common-neighbor count would imply with CCPA. As a result the normalization costs CNC accuracy on six of the eight networks, and it does so most severely on the sparse neural network CEG ($-0.0548$); CNC is ahead only on Jazz ($+0.0056$) and, by $0.0001$, on GrQc, so that CCPA is the stronger of the two in panel average ($0.9126$ versus $0.9017$). Gravity trails DR-RA in panel average with the largest gap (mean $+0.0702$), and CNPop is likewise outperformed in panel average ($+0.0152$); the two per-dataset exceptions are both on wiki-Vote, where Gravity (0.9368) and CNPop (0.9195) overtake DR-RA (0.9099). This result is consistent with the computational structure of the two: as a global method, RWR computes a diffusion similarity for each node pair, and its computational cost (matrix inversion, $O(n^3)$) is of the same order as the closed-form solution of DomiRank; but their roles in the fusion paradigm differ: DomiRank only needs to be solved once offline on the training graph, yielding $n$ node-level scalars, and the centrality product at prediction time is an $O(1)$ table lookup, so the actual online overhead of the fused indices is of the same order as that of local methods. The value of the fusion paradigm therefore lies not in reducing the total computational complexity to the local level, but in condensing global structural information into a reusable node-level representation, trading a one-off offline cost for near-global accuracy.

Beyond AUC, we report Precision@$L$ as defined in Section~3.4 to examine the effect of fusion on the top of the prediction list; the results are shown in Table~\ref{tab:prec}. The Precision of DR-RA is slightly lower than the RA baseline on seven of the eight networks (e.g., 0.2591 versus 0.2773 on BUP), and on FB the two differ by $0.0007$ with DR-RA ahead (0.5302 versus 0.5295); relative to the comparison methods Katz, RWR, CCPA, Gravity and CNPop, DR-RA maintains an advantage on most networks. The gains of our fusion therefore manifest mainly in the global ranking quality characterized by AUC, while the hit rate at the very top of the prediction list is still dominated by local structural information; the role of centrality completion is to improve the global ordering rather than to create new head hits. This observation agrees with the mechanism decomposition of Section~5.6: the completion term only acts on zero-score node pairs without common neighbors, which are usually not at the top of the score sequence.

\begin{table}[!t]
	\centering
	\caption{Precision@$L$ of DR-RA and representative indices ($L$ = number of test edges per fold)}
	\footnotesize
	\setlength{\tabcolsep}{4pt}
	\begin{tabular}{@{}lcccccccc@{}}
		\toprule
		 & \textbf{BUP} & \textbf{CEG} & \textbf{NS} & \textbf{Jazz} & \textbf{Email} & \textbf{FB} & \textbf{GrQc} & \textbf{Wiki} \\
		\midrule
		DR-RA & 0.2591 & 0.1534 & 0.5374 & 0.5420 & 0.3209 & 0.5302 & 0.5535 & 0.1198 \\
		RA & 0.2773 & 0.1543 & 0.5396 & 0.5949 & 0.3404 & 0.5295 & 0.5605 & 0.1218 \\
		Katz & 0.2477 & 0.1478 & 0.3198 & 0.5245 & 0.2681 & 0.3879 & 0.3883 & 0.1468 \\
		RWR & 0.2773 & 0.1538 & 0.4132 & 0.4704 & 0.2276 & 0.2357 & 0.2395 & 0.0229 \\
		CCPA & 0.2477 & 0.1445 & 0.3824 & 0.5321 & 0.2699 & 0.3956 & 0.3964 & 0.1442 \\
		Gravity & 0.1727 & 0.1166 & 0.1000 & 0.1810 & 0.1666 & 0.2075 & 0.1879 & 0.0930 \\
		CNPop & 0.2455 & 0.1604 & 0.3176 & 0.4558 & 0.2278 & 0.3856 & 0.3300 & 0.1389 \\
		\bottomrule
	\end{tabular}
	\label{tab:prec}
\end{table}

\subsection{Parameter sensitivity: the near-critical plateau of the competition intensity \texorpdfstring{$\sigma$}{sigma}}

\begin{figure}[!t]
	\centering
	\includegraphics[width=0.98\textwidth]{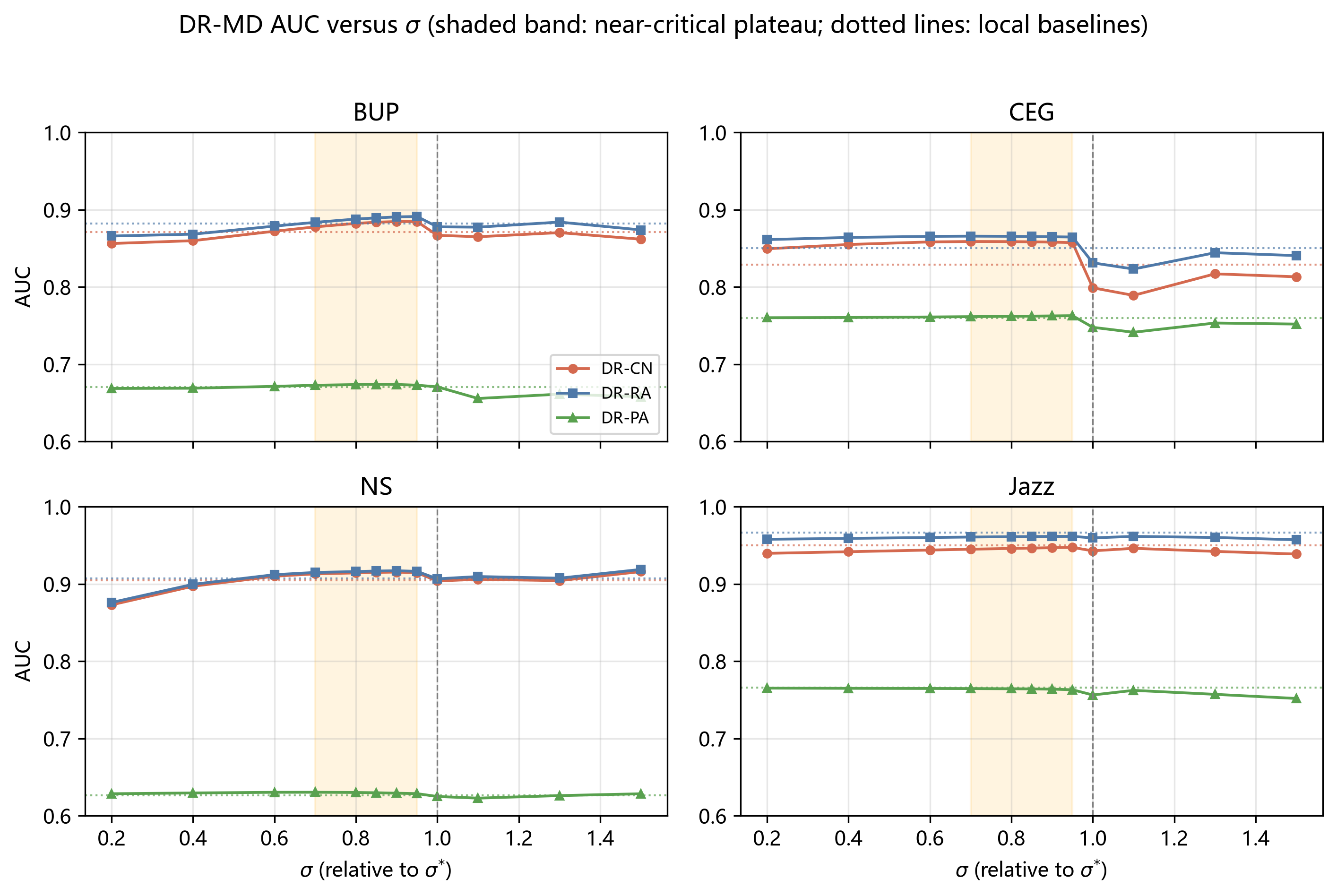}
	\caption{AUC of DR-MD indices versus $\sigma$ (relative to $\sigma^{*}$); the shaded band marks the near-critical plateau and the dotted lines mark the corresponding local baselines. The plateau fluctuations quoted in the text are ranges (maximum minus minimum of the five-fold mean over $\sigma\in\{0.7,0.8,0.85,0.9,0.95\}$), not standard deviations; for DR-CN they are $0.69$, $0.14$, $0.23$ and $0.24$ percentage points on BUP, CEG, NS and Jazz}
	\label{fig:sigma}
\end{figure}

Besides $(\omega,c)$, the competition intensity $\sigma$ of DomiRank itself is the third parameter of the fused indices. Figure~\ref{fig:sigma} shows the AUC variation of DR-CN, DR-RA and DR-PA on four datasets for $\sigma\in[0.2,1.5]$ (relative to $\sigma^{*}$), with $(\omega,c)$ fixed at Eq.~\eqref{eq:params}. Three regularities can be observed. (1) \textbf{Near-critical plateau}: within $\sigma\in[0.7,0.95]$, the AUCs of all indices enter a stationary peak region, and the within-plateau fluctuations of DR-CN on the four datasets do not exceed 0.70 (BUP), 0.14 (CEG), 0.23 (NS) and 0.24 (Jazz) percentage points, respectively; the $\sigma=0.85$ adopted in this paper lies exactly inside this plateau. (2) \textbf{Super-critical degradation}: for $\sigma\geq 1$, the matrix $(\sigma A+I)$ approaches singularity and the performance drops markedly, e.g., DR-CN on CEG falls from 0.858 to 0.799 ($-5.9$ percentage points), in agreement with the theoretical analysis of Engsig et al.~\cite{A6} on the effective domain $(0,\sigma^{*})$ of DomiRank. (3) \textbf{Sub-critical failure}: when $\sigma$ is too small, the fusion may instead fall below the baseline, e.g., on BUP at $\sigma=0.2$, DR-CN is 0.8563, below the CN baseline 0.8709; in this regime $\mathrm{DR}(\cdot)$ degenerates into an approximate degree index, the product $S^{DR}$ becomes highly collinear with the degree product, and the modulation term loses its discriminative power. Together, the three regularities show that $\sigma$ is not ``the larger the better'': the near-critical region $[0.7,0.95]\sigma^{*}$ is a safe selection interval combining performance and robustness.

\subsection{Mechanism decomposition: differentiated contributions of completion and modulation}

The fusion rule~\eqref{eq:drfusion} contains two mechanisms: ``multiplicative modulation'' (the $\omega$ term) and ``zero-score completion'' (the $c$ term). To decompose their contributions, we set $\omega=0$ with $c=0.1$ fixed (switching off modulation and keeping completion only) and compare against the full model ($\omega=5$, $c=0.1$); the results are shown in Table~\ref{tab:decomp}.

\begin{table}[!t]
	\centering
	\caption{Mechanism decomposition: completion-only versus full fusion against local baselines (fold-level tests, 40 pairs)}
	\footnotesize
	\setlength{\tabcolsep}{6pt}
	\begin{tabular}{@{}lcccl@{}}
		\toprule
		Index & Configuration & $\Delta$AUC & Wins/40 & $p$-value \\
		\midrule
		CN & $\omega=0.0$ & $+0.0049$ & 31/40 & $1.2\times10^{-4}$ \\
		CN & $\omega=5.0$ & $+0.0077$ & 29/40 & $2.3\times10^{-4}$ \\
		AA & $\omega=0.0$ & $+0.0049$ & 31/40 & $1.2\times10^{-4}$ \\
		AA & $\omega=5.0$ & $+0.0047$ & 29/40 & $1.7\times10^{-3}$ \\
		RA & $\omega=0.0$ & $+0.0046$ & 31/40 & $2.7\times10^{-4}$ \\
		RA & $\omega=5.0$ & $+0.0042$ & 29/40 & $3.3\times10^{-3}$ \\
		JC & $\omega=0.0$ & $+0.0048$ & 31/40 & $1.2\times10^{-4}$ \\
		JC & $\omega=5.0$ & $+0.0091$ & 34/40 & $3.5\times10^{-6}$ \\
		HPI & $\omega=0.0$ & $+0.0049$ & 31/40 & $1.2\times10^{-4}$ \\
		HPI & $\omega=5.0$ & $+0.0082$ & 34/40 & $5.6\times10^{-6}$ \\
		SO & $\omega=0.0$ & $+0.0049$ & 31/40 & $1.2\times10^{-4}$ \\
		SO & $\omega=5.0$ & $+0.0094$ & 34/40 & $2.3\times10^{-6}$ \\
		PA & $\omega=0.0$ & $0.0000$ & 0/40 & --- \\
		PA & $\omega=5.0$ & $+0.0011$ & 28/40 & $1.5\times10^{-3}$ \\
		\bottomrule
	\end{tabular}
	\label{tab:decomp}
\end{table}

Table~\ref{tab:decomp} reveals clear structural regularities. (1) \textbf{The completion mechanism is a universal gain base}. With only the small-dose completion $c=0.1$ retained, all six common-neighbor-dependent indices obtain a consistent significant improvement of about $+0.0046\sim +0.0049$ ($p\leq 2.7\times10^{-4}$, 31/40 pairs positive); the value is almost identical across indices because the completion score $c\cdot S^{DR}$ of zero-score node pairs is independent of the form of the local index; PA, having no zero branch, is completely unaffected by pure completion (0/40), which validates the control logic. (2) \textbf{The contribution of the modulation mechanism varies by index family}: for the normalized indices JC/SO/HPI, adding modulation nearly doubles the gain ($+0.0048\sim +0.0049\to +0.0082\sim +0.0094$) and raises the win rate from 31/40 to 34/40, showing that multiplicative modulation provides an effective reordering of the with-common-neighbor pairs of these indices; for the count-based indices AA/RA, modulation brings a slight loss of about $-0.0004\sim -0.0001$ (their optimal $\omega$ is close to 0.5, at the left end of the plateau of regularity (2) in Section~5.2); for CN it contributes about $+0.0028$; for PA, the $+0.0011$ gain of the full fusion comes entirely from the modulation term (28/40 pairs positive); PA itself already carries global information in the form of a degree product and is unaffected by completion, but multiplicative modulation still brings a small reordering benefit. (3) The two mechanisms are \textbf{complementary} under the unified parameters: completion guarantees that ``no index is harmed'', modulation provides the main gains for the normalized indices, and together they fully explain the consistent significant improvements of all seven indices.

\subsection{Degree-corrected benchmark: which gains survive}
\label{sec:degcorr}

Aiyappa et al.~\cite{A46} showed that the standard link-prediction benchmark carries an implicit degree bias: because positive samples are drawn from real edges, their endpoints are degree-biased, and a score built from node degrees alone attains an average AUC-ROC of $0.83$ over $90$ graphs, ranking thirteenth among the twenty-six methods compared. Their Algorithm~1 removes this channel by drawing each negative sample from a degree-weighted node table, so that positive and negative pairs are matched in degree. Two later studies confirm that the channel is a property of the evaluation protocol rather than of any particular scoring design: Wang and Derr~\cite{A52} show that link prediction performance differs systematically across degree classes and that the choice of evaluation metric interacts with this imbalance, and Y{\i}lmaz et al.~\cite{A51} demonstrate that uniformly sampled test sets bias link prediction evaluation towards high-degree nodes, a bias that persists even across network snapshots, and propose bias-aware alternatives. Because several of the quantities studied here contain popularity-type information (PA is a degree product, and the DomiRank score is itself correlated with degree at $\rho=0.34$--$0.62$, Section~6.1), we repeat the comparison under this degree-corrected protocol. The same logic, namely separating what a degree baseline already explains from what remains, is also being pursued by the authors for attack fragility, where the configuration-model robustness plays the role of the degree-only reference; that companion study is currently under review and no conclusion of the present paper depends on it~\cite{A48}.

The experiment is designed so that the only quantity that changes is the negative-sample set. Each network is evaluated in three arms computed in the same pass, sharing the fold partition (seed $1000+k$), the positive samples and the scoring functions, in particular the unified fusion parameters $\omega=5$ and $c=0.1$: arm~$A$ uses all non-edges of the training graph as negatives, the protocol behind Tables~\ref{tab:main} and~\ref{tab:wilcox}; arm~$A'$ draws $1{:}1$ negatives uniformly from that same pool, aligning the sample size with arm~$B$; arm~$B$ draws $1{:}1$ negatives with the degree-corrected sampling of Algorithm~1. Arm~$A$ reproduces the archived main table to within $6.8\times10^{-7}$ in absolute AUC over all methods and networks, which confirms that the pipeline is unchanged and that the differences below are effects of the negative-sample distribution rather than of a different code path. That maximum is a per-network, per-method quantity and is therefore not visible in any panel average; the reader-facing counterpart of this fidelity check is the first data column of Table~\ref{tab:degcorr_gain}, which reproduces the mean gain and the positive-pair count of the first block of Table~\ref{tab:wilcox} row for row ($+0.0077$/29 for CN through $+0.0011$/28 for PA, identical in all fourteen cells). Both implementation degrees of freedom left open by Algorithm~1 were varied: the degree table is built either from the full graph or from the training graph, and each positive pair is matched with either $1$ or $10$ negatives (the default of the official implementation). None of the conclusions below depends on this choice.

Table~\ref{tab:degcorr} reports the panel-average AUCs and Table~\ref{tab:degcorr_gain} the gains, and three points stand out. (i) \textbf{The degree channel is closed.} The degree-only baseline collapses from $0.7736$ to $0.5078$ in panel average, that is, to chance level, with the largest drop on wiki-Vote ($-0.402$). This is the phenomenon reported by Aiyappa et al. ($0.83\to0.54$) reproduced on this panel, and it certifies that the degree correction has actually taken effect. (ii) \textbf{The gains over the local baselines do not survive.} The seven gains move from $+0.0011\sim+0.0094$ under arm~$A$ to $-0.0117\sim+0.0006$ under arm~$B$. Pooling the $8\times5=40$ fold-level pairs and testing $\Delta$(DR-$m$ $-$ $m$) one-sided, the arm-$A$ gains are all significant ($p=2.3\times10^{-6}\sim3.3\times10^{-3}$, with 28--34 of 40 pairs positive), whereas under arm~$B$ every comparison except DR-PA reverses sign ($p>0.9997$, with 1--12 of 40 pairs positive); in particular DR-RA falls below RA on all eight networks (Table~\ref{tab:degcorr_pernet}; per-network margins $-0.0343$ on GrQc down to $-0.0004$ on FB). The change is one of sign rather than of magnitude, and it appears in every index family. Two features of Table~\ref{tab:degcorr_gain} should be read together, since they are what licenses the interpretation above: the sign of the arm-$B$ $\Delta$AUC column and the Pos./40 count move together, a comparison reversing sign exactly when its positive-pair count falls below half of 40, which is the case for all six reversed rows; and DR-PA is the single exception on both readings, its gain staying positive but marginal with 22 of 40 pairs positive. The sign reversal of the other six indices is therefore not an artefact of the test but a property of the majority of pairs. The $p$ values of the arm-$B$ column belong to a separate hypothesis family of seven tests and are not BH-corrected; with every one of them at $p>0.9997$ no correction could alter the conclusion, but the two families should not be compared term by term. No arm-$B$ contrast attains $p=1$ exactly, each having at least one positive pair. (iii) \textbf{What does survive is the lead over the degree-only reference.} The margin of DR-RA over PA widens from $+0.1520$ to $+0.3341$ in panel average, positive in $40$ of $40$ fold-level pairs ($p=9.1\times10^{-13}$).

The reading is consistent with the decomposition of Section~5.6. The completion term $c\cdot S^{DR}$ gives a score to zero-score pairs, those without common neighbors, that increases with node importance; under exhaustive negative sampling, high-importance pairs without common neighbors are precisely the negatives that most resemble true edges, and lifting them is where much of the gain comes from. Under the degree-corrected protocol the same lift no longer separates positives from negatives, and the completion term degenerates into an almost constant factor. This is not in tension with Section~6.1, which already reports that the DomiRank score correlates with degree at $\rho=0.34$--$0.62$: the degree-corrected benchmark switches that channel off, leaving the common-neighbor information, which is exactly where local indices are already strong.

Two consequences should be stated plainly. First, the statement of Section~5.3 that all seven fused indices improve significantly over their local baselines is a statement about the standard protocol and is limited to it. Second, the statement that survives the degree-corrected protocol is the comparison with a degree-only reference: on a benchmark from which degree information has been removed, DR-RA stays far above what the degree sequence alone can achieve, while it no longer leads the local baselines: under arm~$B$ the panel average of DR-RA ($0.8419$) falls below that of RA ($0.8534$), and Table~\ref{tab:degcorr} reports only these three indices, so no ranking among the global or advanced baselines is claimed under the degree-corrected protocol. Both observations are consistent with the design principle of Section~4: fusion transfers global structural information into a local score, and the advantage is largest exactly where degree information is legitimately available and the local score is blind.

\begin{table}[!t]
	\centering
	\caption{Panel-average AUC over the eight networks under arm $A$ (exhaustive negatives, protocol of Tables~\ref{tab:main} and~\ref{tab:wilcox}), arm $A'$ (uniform $1{:}1$ negatives) and arm $B$ (degree-corrected $1{:}1$ negatives, Algorithm~1 of Ref.~\cite{A46})}
	\footnotesize
	\setlength{\tabcolsep}{8pt}
	\begin{tabular}{@{}lccc@{}}
		\toprule
		Index & Arm $A$ & Arm $A'$ & Arm $B$ \\
		\midrule
		PA (degree only) & 0.7736 & 0.7725 & \textbf{0.5078} \\
		RA & 0.9215 & 0.9218 & 0.8534 \\
		DR-RA & 0.9257 & 0.9256 & 0.8419 \\
		\bottomrule
	\end{tabular}
	\label{tab:degcorr}
\end{table}

\begin{table}[!t]
	\centering
	\caption{Per-network AUC under arm $B$ (degree-corrected $1{:}1$ negatives) for the two indices of Table~\ref{tab:degcorr} that are compared there. DR-RA is below RA on all eight networks, which is the per-network basis of the corresponding statement in Section~\ref{sec:degcorr}}
	\footnotesize
	\setlength{\tabcolsep}{6pt}
	\begin{tabular}{@{}lrr@{}}
		\toprule
		Network & DR-RA & RA \\
		\midrule
		BUP   & 0.8271 & 0.8388 \\
		CEG   & 0.7519 & 0.7532 \\
		NS    & 0.8899 & 0.9048 \\
		Jazz  & 0.9165 & 0.9342 \\
		Email & 0.8137 & 0.8189 \\
		FB    & 0.9831 & 0.9835 \\
		GrQc  & 0.8707 & 0.9050 \\
		Wiki  & 0.6824 & 0.6884 \\
		\midrule
		Panel average & 0.8419 & 0.8534 \\
		\bottomrule
	\end{tabular}
	\label{tab:degcorr_pernet}
\end{table}

\begin{table}[!t]
	\centering
	\caption{Gains of the fused indices over their local baselines under the two protocols, pooled over the $8\times5=40$ fold-level pairs: one-sided Wilcoxon signed-rank test (alternative: the gain is positive) and the number of positive pairs. Under arm $B$ every comparison except DR-PA is non-significant with $p>0.9997$; the smallest values are attained by JC and SO ($p=0.9998$). DR-PA is the single exception on both readings: its gain stays positive but marginal ($+0.0006$) with 22 of 40 pairs positive, and it is the only comparison whose arm-$B$ $p$ is not close to one ($p=0.16$).}
	\footnotesize
	\setlength{\tabcolsep}{5pt}
	\begin{tabular}{@{}lcclcc@{}}
		\toprule
		 & \multicolumn{3}{c}{Arm $A$ (exhaustive)} & \multicolumn{2}{c}{Arm $B$ (degree-corrected)} \\
		\cmidrule(lr){2-4}\cmidrule(lr){5-6}
		Pair & $\Delta$AUC & $p$ & Pos./40 & $\Delta$AUC & Pos./40 \\
		\midrule
		DR-CN $-$ CN & $+0.0077$ & $2.3\times10^{-4}$ & 29 & $-0.0100$ & 5 \\
		DR-AA $-$ AA & $+0.0047$ & $1.7\times10^{-3}$ & 29 & $-0.0117$ & 1 \\
		DR-RA $-$ RA & $+0.0042$ & $3.3\times10^{-3}$ & 29 & $-0.0114$ & 1 \\
		DR-JC $-$ JC & $+0.0091$ & $3.5\times10^{-6}$ & 34 & $-0.0069$ & 11 \\
		DR-HPI $-$ HPI & $+0.0082$ & $5.6\times10^{-6}$ & 34 & $-0.0102$ & 3 \\
		DR-SO $-$ SO & $+0.0094$ & $2.3\times10^{-6}$ & 34 & $-0.0073$ & 12 \\
		DR-PA $-$ PA & $+0.0011$ & $1.5\times10^{-3}$ & 28 & $+0.0006$ & 22 \\
		\bottomrule
	\end{tabular}
	\label{tab:degcorr_gain}
\end{table}

\section{Discussion}

\subsection{Why unnormalized PageRank fusion fails: scale calibration, information content and the dual-mechanism division of labor}

The controlled experiments of Sections~4 and~5 support the following reading: scale calibration and information content jointly determine fusion effectiveness. On the scale side: the PageRank product has scale $O(n^{-2})$ and its multiplicative factor is nearly identity, and the grid check of Section~4.2 shows that the modulation weight changes the average gain by less than $10^{-6}$ across a $24\times$ range of $\omega$, i.e.\ nothing at any reportable precision; The control experiments of Section~4.4 qualify this further. Even when the PR product is calibrated by order-preserving scaling to a magnitude comparable with that of DomiRank, its gain is still only $1/3$ to $1/9$ of that of the DR arm, and none of the six common-neighbor indices is significant. On the information side: the rank-correlation diagnostic of Section~4.2, which reports the per-network values behind the interval $0.899\sim0.990$ between PR and degree against $0.341\sim0.621$ for DomiRank, and the control arm that replaces centrality by node degree closely resembles the PR min-max arm Used as a predictor on its own, the centrality product gives full-candidate-set AUCs of $0.6596$ for the normalized degree product against $0.6553$ for the PR min-max arm on BUP, and $0.6268$ against $0.6256$ on NS, the largest difference occurring on Jazz at $0.0107$. The scale-calibrated PR product therefore still carries degree information, and the degree product is precisely the information captured by the weakest PA baseline of this paper. This yields a transferable two-condition criterion: \textbf{a centrality index, before entering the fusion framework, must satisfy at the same time (i) that the scale of its product is comparable to that of the local scores, and (ii) that its product is not strongly collinear with the local indices.} Normalization is not an inconsequential preprocessing step but an inseparable part of the fusion rule. The normalization scheme is itself a variable, since min-max is an affine map with a shift that alters the ordering of the node product (Section~4.4 gives direct evidence on BUP), so that an order-preserving scaling such as $C/\max C$ should be used when the design intent is only to calibrate the scale. Under this criterion, other centralities such as k-shell, HITS and betweenness all have the potential to become effective fusion carriers, whereas any centrality that is not scale-calibrated, or that is highly redundant with the local information, may repeat the failure of PageRank. An additional advantage of DomiRank is that its $\sigma$ parameter allows fine-tuning the ``global--local'' information scale within the near-critical plateau, so as to adapt to networks of different densities and clustering structures (Section~5.5). The conclusions of this subsection concern the \emph{unnormalized} PageRank scores of the original Charikhi protocol, and are not a general assessment of the PageRank centrality itself: the value of PageRank in web ranking, influence analysis and random-walk modelling is unaffected by the present results.

The cross-comparison on wiki-Vote (Sections~4.3 and~5.3) extends the above two-condition criterion into a more complete \textbf{dual-mechanism division of labor} framework. The two mechanisms of the fusion rule~\eqref{eq:fusion} are governed by different factors and their doses should be designed separately. (i) The gain of the \textbf{completion mechanism} is jointly determined by the fraction of zero-score node pairs and the standalone predictive power of the centrality product. Wiki possesses both a high zero-score fraction (0.89) and the strongest PA baseline (0.9332), and PR completion obtains the largest gain in this paper ($+0.036$, attributed to the completion mechanism in Section~4.3); NS/GrQc have even higher zero-score fractions but weak PA baselines, and the same completion yields almost no benefit. This indicates that the completion dose $c$ should be chosen \textit{network by network}: it can be larger on sparse networks where the centrality product is itself predictive, and should tend to zero otherwise. (ii) The effectiveness of the \textbf{modulation mechanism} is determined jointly by the dynamic range of the centrality product and its information redundancy. The PR product is nearly identity, so its modulation is ineffective but harmless; the DR product has a moderate range, so its modulation is effective on most networks but disturbs the ordering on near-tree-like networks where centrality correlates weakly with local ordering (on Wiki, all seven DR-MD indices fall below the baselines, net $-0.007$). The two mechanisms can point in opposite directions on the same network: on Wiki, PR fusion (identity modulation $+$ effective completion) is significantly better than DR fusion (harmful modulation $+$ effective completion), showing that ``which centrality to choose'' and ``which mechanism to enable, at what dose'' are two independent design decisions. The resulting practical rule is: \textit{first assess the standalone predictive power of the centrality product on the target network with cheap diagnostics such as the PA baseline to decide the completion dose; then check whether the modulation mechanism takes effect via the dynamic range to decide the modulation weight}.

\subsection{Limitations}

This paper has the following limitations. (1) \textbf{The network scale still has an upper bound}. Our datasets span two orders of magnitude in node size, from 105 to 7066, with three networks exceeding $10^3$ nodes; we have not tested networks of $10^4$ nodes or more. Although Engsig et al.~\cite{A6} have shown that DomiRank is parallelizable and applicable to massive networks, and all our experiments were completed with sparse solvers, the consistency of the fusion conclusions and the computational scalability on larger networks remain to be systematically verified; the consistency of conclusions on more typical benchmark networks also awaits supplementation. (2) \textbf{The unified parameters are selected and evaluated on the same fold partitions}, so the reported values carry a mild selection bias; although Section~5.2 provides three layers of mitigating evidence (plateau property, mechanism decomposition and leave-one-network-out validation) and the $10\times5$ repeated cross-validation of Section~5.3 further confirms robustness to fold-partition randomness, fully nested cross-validation remains a more rigorous confirmation scheme; moreover, the leave-one-network-out validation of Section~5.2 uses a different selection criterion (mean $\Delta$AUC instead of maximin), a coarser grid, and only the representative index DR-RA, so its independence is limited: its role is to rule out catastrophic overfitting of the parameter choice rather than to estimate generalization performance precisely, and it should not be counted as a third independent confirmation. (3) \textbf{The evaluation protocol, though fair, is single}. We fix exhaustive negative-sample comparison and a 20\% probe-edge ratio; performance under other protocols (e.g., 10\% hold-out, balanced negative sampling, temporal splitting) remains to be systematically examined~\cite{A45}. (4) \textbf{Gains narrow on dense and large-scale networks}. As described in Section~5.3, on Jazz and Wiki some or even all DR-MD indices fail to exceed the baselines; the applicable scenario of the fusion strategy is networks of medium or lower density and medium scale. The cross-comparison on Wiki, where PR fusion obtains the largest gain through the completion mechanism, further suggests that the choice of fusion carrier (centrality index) and of enabled mechanisms should vary with network topology (Section~6.1). (5) The magnitude of the gains is at the permille-to-percent level, a stable but moderate improvement; on datasets where AUC is nearly saturated (e.g., the local baseline on FB already reaches 0.99, and 0.90 on NS), the room for further absolute gains is limited. (6) \textbf{The protocol, though fair, retains a degree channel}, in the sense of Ref.~\cite{A46}. Because negative samples are drawn from non-edges of the observed graph, a score built from node degrees alone still receives a substantial share of the credit for the observed ranking. Section~5.7 quantifies this: under a degree-corrected benchmark the gains over the local baselines vanish and reverse in sign, while the lead over a degree-only reference is preserved. A benchmark that removes the channel completely (a degree-matched analysis, or a temporal split in which scores cannot be tuned on the same snapshot) would give a sharper separation between the structural and the popularity parts of the improvement. (7) \textbf{Two quantitative claims are stated as scope limits rather than as measured results}. First, the attribution of the wiki-Vote gain in Section~4.3 to the completion mechanism is established by the internal control reported there (the completion-free PA index gains $+5\times10^{-6}$ on that network) together with the near-identity scale analysis of Section~4.2; the completion-only arm ($\omega=0$) has not been evaluated on wiki-Vote in isolation, and the completion-only magnitude reported in Section~5.6 ($+0.0046\sim +0.0049$) is pooled over the eight networks, so it should not be read as the wiki-Vote value. Second, the baseline comparisons of Section~5.4 are made at the parameter values reported in the original literature and are not tuned on our evaluation partitions; the resulting bias is one-sided and can only favour DR-RA, but its magnitude is not quantified here, because that would require a per-fold sweep of the four closest baselines (Katz, RWR, CN2D, CNC) whose results are then not directly comparable at a single parameter value. (8) \textbf{The normalization scheme assumes global knowledge.} The min-max map of Eq.~\eqref{eq:minmax} reads off the maximum and the minimum of the centrality over the whole graph, so both ends must be recomputed whenever the node set changes, and a single newly added extreme node rescales every other node's value. This is harmless for the static benchmarks studied here but blocks the direct transfer of the fusion rule to incremental or streaming settings, where an order-preserving map such as $C/\max C$ (Section~4.4) or a running extremum would be required instead.

\section{Conclusion and outlook}

\subsection{Conclusion}

Within a unified piecewise fusion framework (multiplicative modulation $+$ small-dose completion), this paper systematically studies the ``centrality$\,\times\,$local similarity'' link prediction paradigm under a fair protocol, and arrives at five main conclusions. (1) The scale calibration and the information content of the centrality product jointly determine whether the modulation mechanism works: as the scale analysis predicts and the experiments confirm, the multiplicative factor of the PageRank product ($O(n^{-2})$) is a near-identity map and the product is highly redundant with node degree (rank correlation $0.899\sim0.990$), so that its fusion brings no consistent cross-network statistically significant improvement for six of the seven local indices (the only significant gain, for PA, is of very small magnitude); the control experiments on scale treatment further show that normalization is a necessary but not a sufficient condition, and that the normalization scheme itself affects the outcome, order-preserving scaling being superior to min-max, the latter altering the ordering of the node product; the single exception is the uniform gain of about $+0.036$ on wiki-Vote, which Section~4.3 attributes entirely to the completion mechanism and Section~6.1 reads as a joint consequence of the zero-score fraction and the standalone predictive power of the centrality product. (2) The min-max normalized DomiRank product has an $O(1)$ dynamic range; under unified parameters ($\omega=5$, $c=0.1$), the normalized fused indices DR-JC/HPI/SO significantly outperform their local baselines already at the dataset level (Wilcoxon and sign tests $p=0.027$ and $0.035$ on 8 independent networks), and all seven DR fused indices achieve fold-level significant improvements ($p\leq3.3\times10^{-3}$, BH-corrected $q\leq4.9\times10^{-3}$), the gains of CN/AA/RA/PA being direction-consistent but not reaching dataset-level significance; the DR indices are significantly better than the corresponding PR fused indices on CN/JC/HPI/SO; a $10\times5$ repeated cross-validation confirms that the main conclusions are robust to fold-partition randomness. Under the ranking-based Precision@$L$ metric the picture is more conservative: DR-RA slightly trails its RA baseline on seven of the eight networks, indicating that the fusion gain is concentrated in the global ordering measured by AUC rather than in the very top of the ranking, a trade-off to be weighed in applications where only the top-ranked predictions are acted upon. (3) DR-RA attains an average AUC of 0.9257, outperforming nine of the ten comparison methods (Katz, SimRank, LNB, CN2D, CNC, CND, CCPA, Gravity and CNPop) and being comparable to the global method RWR (mean $-0.0003$); since the DomiRank centrality is computed only once offline on the training graph and the prediction stage requires only $O(1)$ table lookups and local operations, the fused indices achieve global-level accuracy while keeping the online overhead close to that of local methods; the goal of the paradigm is not to reduce the total computational complexity (the centrality solve remains a one-off offline cost and is parallelizable~\cite{A6}), but to condense global structural information into a reusable node-level scalar representation. (4) Parameter design follows three quantitative principles: the completion coefficient is constrained by a safety boundary ($0<c\leq0.3$), the modulation weight exhibits a wide plateau ($\omega\in[0.5,12]$), and the competition intensity should be chosen in the near-critical region ($\sigma\in[0.7,0.95]\sigma^{*}$); mechanism decomposition and the wiki-Vote cross-comparison further show that the completion gain is determined by the zero-score fraction and the standalone predictive power of the centrality product, while the effectiveness of modulation is determined by the dynamic range, so the doses of the two mechanisms should be designed separately. These results provide a reproducible protocol benchmark, explicit applicability boundaries and transferable design principles for the fusion paradigm. (5) The improvement reported here is specific to the standard protocol: under a degree-corrected benchmark that matches positive and negative pairs in degree, the gains over the local baselines disappear and reverse in sign, whereas the lead of DR-RA over a degree-only score is preserved and widens (from $+0.1520$ to $+0.3341$ in panel average, positive in $40/40$ fold-level pairs), so it is the comparison against a degree-only reference, rather than the margin over the local baseline, that survives when the popularity channel is closed.

\subsection{Outlook}

The core idea of our framework is to regard any node centrality index $C(\cdot)$, after normalization, as a ``scalarized representation of global importance'', and to provide a global score for a node pair through $C(x)\cdot C(y)$. Beyond PageRank and DomiRank, k-shell decomposition~\cite{A32} (low cost, near-linear complexity, suitable for large-scale networks), HITS~\cite{A33} (Hub/Authority dual centrality, suitable for directed networks; Chen et al.~\cite{A38} have provided independent empirical evidence in directed scenarios), betweenness centrality~\cite{A34} (sensitive to bridging structures), eigenvector centrality~\cite{A35} and GNN-based learned centralities~\cite{A29,A31} can all be instantiated under our design principles (normalize first, then fuse with unified parameters); the centrality fusion in hypernetwork scenarios of Nandini et al.~\cite{A41} and the popularity--similarity decomposition of He et al.~\cite{A40} also provide concurrent support for this direction. Several open problems remain at the methodological level: (1) an adaptive learning mechanism for the fusion weights $(\omega,c)$, upgrading ``grid-search selection'' to ``supervised optimization on the training set'', together with nested cross-validation to eliminate selection bias; (2) generalization of the fusion paradigm to directed, weighted and dynamic temporal networks; (3) approximation algorithms and scalability verification on large-scale networks ($|V|>10^4$); (4) normalization schemes for local similarity scores and a quantitative analysis of the theoretical optimality of fusion parameters. The protocol benchmark and the design principles established here are offered as a step toward higher and more credible prediction accuracy at low computational cost for the ``centrality--local similarity'' method family.

\section*{Data availability}
The complete replication package, comprising the evaluation-protocol code, the derived
result tables, and the locked inputs for the eight real-world networks, is
openly available in the Zenodo repository at
\url{https://doi.org/10.5281/zenodo.23008698}. This is a concept DOI, which
resolves to the latest version of the record; readers reproducing the
tables of Sections 4--5 should therefore select the deposited version whose
release matches this manuscript rather than the latest one.

\section*{Version and protocol evolution}
An earlier version of this manuscript was posted to arXiv as v1. It reported the
same fusion family on a panel of nine networks, and concluded that the fused
indices improved consistently on their local baselines on all nine of them
($p=0.002$, with a mean AUC of $0.7084$ for DR-RA). Three protocol changes
separate the present version from it, each adopted to remove a source of bias
rather than to improve the reported outcome. The panel was replaced: all nine
networks of v1 had at most $500$ nodes, whereas the present panel spans $105$
to $7066$ nodes and four of the eight networks do not appear in v1 at all.
Negative samples are now enumerated exhaustively. v1 drew them at ratios of
$1{:}1$, $1{:}5$ and $1{:}10$ to the positive set and argued from the
paired-ranking definition of AUC that uniform subsampling leaves the expectation
unchanged; that holds for the expectation of AUC but not for a comparison across
indices, since a sampled candidate set admits score-based filtering, the very
shortcut removed in Section~4.1. The parameters of the comparison methods are
now stated in Section~5.4 together with the direction of the bias they leave,
which v1 did not discuss. Under these conditions the reported improvement does
not survive for six of the seven local indices, and DR-RA reaches a mean AUC of
$0.9257$. The two sets of numbers are not comparable and are not a
before-and-after record of a fixed quantity: they differ simultaneously in
panel, negative-sample construction, baselines and hypothesis set, and the two
AUC figures average over different network sets. Only the present version
implements the protocol of Section~3.3.



\begin{thebibliography}{10}

\bibitem{A8}
David Liben-Nowell and Jon Kleinberg.
\newblock The link-prediction problem for social networks.
\newblock {\em Journal of the American Society for Information Science and
  Technology}, 58(7):1019--1031, 2007.

\bibitem{A14}
Hanghang Tong, Christos Faloutsos, and Jia-Yu Pan.
\newblock Random walk with restart: fast solutions and applications.
\newblock {\em Knowledge \& Information Systems}, 14(3):327--346, 2008.

\bibitem{A16}
Weiping Liu and Linyuan L{\"u}.
\newblock Link prediction based on local random walk.
\newblock {\em EPL (Europhysics Letters)}, 89(5):58007, 2010.

\bibitem{A12}
Paul Jaccard.
\newblock {\'E}tude comparative de la distribution florale dans une portion des
  alpes et des jura.
\newblock {\em Bulletin de la Soci{\'e}t{\'e} Vaudoise des Sciences
  Naturelles}, 37:547--579, 1901.

\bibitem{A9}
Lada~A. Adamic and Eytan Adar.
\newblock Friends and neighbors on the web.
\newblock {\em Social Networks}, 25(3):211--230, 2003.

\bibitem{A10}
Tao Zhou, Linyuan L{\"u}, and Yi-Cheng Zhang.
\newblock Predicting missing links via local information.
\newblock {\em European Physical Journal B}, 71(4):623--630, 2009.

\bibitem{A11}
M.~E.~J. Newman.
\newblock Clustering and preferential attachment in growing networks.
\newblock {\em Physical Review E}, 64(2):025102, 2001.

\bibitem{A26}
E.~Ravasz, A.~L. Somera, D.~A. Mongru, Z.~N. Oltvai, and A.-L. Barab{\'a}si.
\newblock Hierarchical organization of modularity in metabolic networks.
\newblock {\em Science}, 297(5586):1551--1555, 2002.

\bibitem{A27}
T.~S{\o}rensen.
\newblock A method of establishing groups of equal amplitude in plant sociology
  based on similarity of species content and its application to analyses of the
  vegetation on danish commons.
\newblock {\em Biologiske Skrifter}, 5:1--34, 1948.

\bibitem{A13}
Leo Katz.
\newblock A new status index derived from sociometric analysis.
\newblock {\em Psychometrika}, 18(1):39--43, 1953.

\bibitem{A15}
Glen Jeh and Jennifer Widom.
\newblock Simrank: A measure of structural-context similarity.
\newblock In {\em Proceedings of the Eighth ACM SIGKDD International Conference
  on Knowledge Discovery \& Data Mining}, pages 538--543, 2002.

\bibitem{A17}
Linyuan L{\"u}, Ci-Hang Jin, and Tao Zhou.
\newblock Similarity index based on local paths for link prediction of complex
  networks.
\newblock {\em Physical Review E}, 80(4):046122, 2009.

\bibitem{A19}
Iftikhar Ahmad, Muhammad~Usman Akhtar, Salma Noor, and Ambreen Shahnaz.
\newblock Missing link prediction using common neighbor and centrality based
  parameterized algorithm.
\newblock {\em Scientific Reports}, 10:364, 2020.

\bibitem{A18}
Jinxuan Yang and Xiao-Dong Zhang.
\newblock Predicting missing links in complex networks based on common
  neighbors and distance.
\newblock {\em Scientific Reports}, 6:38208, 2016.

\bibitem{A20}
Zhen Liu, Qian-Ming Zhang, Linyuan L{\"u}, and Tao Zhou.
\newblock Link prediction in complex networks: A local na{\"\i}ve bayes model.
\newblock {\em EPL (Europhysics Letters)}, 100(4):48003, 2012.

\bibitem{A25}
Fei Tan, Yongxiang Xia, and Boyao Zhu.
\newblock Link prediction in complex networks: A mutual information
  perspective.
\newblock {\em PLoS ONE}, 9(9):e107056, 2014.

\bibitem{A21}
Ke-ke Shang, Tong-Chen Li, Michael Small, David Burton, and Yan Wang.
\newblock Link prediction for tree-like networks.
\newblock {\em Chaos}, 29(6):061103, 2019.

\bibitem{A23}
Carlo~Vittorio Cannistraci, Gregorio Alanis-Lobato, and Timothy Ravasi.
\newblock From link-prediction in brain connectomes and protein interactomes to
  the local-community-paradigm in complex networks.
\newblock {\em Scientific Reports}, 3:1613, 2013.

\bibitem{A22}
Diyawu Mumin, Lei-Lei Shi, and Lu~Liu.
\newblock An efficient algorithm for link prediction based on local
  information: Considering the effect of node degree.
\newblock {\em Concurrency and Computation: Practice and Experience},
  34(7):e6289, 2022.

\bibitem{A24}
Abdul Samad, Mamoona Qadir, and Ishrat Nawaz.
\newblock Sam: A similarity measure for link prediction in social network.
\newblock In {\em 2019 13th International Conference on Mathematics, Actuarial
  Science, Computer Science and Statistics (MACS)}, pages 1--9, 2019.

\bibitem{A28}
Puneet Kapoor, Sakshi Kaushal, Harish Kumar, and Kushal Kanwar.
\newblock A survey on feature extraction and learning techniques for link
  prediction in homogeneous and heterogeneous complex networks.
\newblock {\em Artificial Intelligence Review}, 57(12):348, 2024.

\bibitem{A30}
Juanhui Li, Harry Shomer, Haitao Mao, Shenglai Zeng, Yao Ma, Neil Shah, Jiliang
  Tang, and Dawei Yin.
\newblock Evaluating graph neural networks for link prediction: Current
  pitfalls and new benchmarking.
\newblock In {\em Advances in Neural Information Processing Systems 36
  (NeurIPS)}, 2023.

\bibitem{A3}
Mourad Charikhi.
\newblock Association of the pagerank algorithm with similarity-based methods
  for link prediction in complex networks.
\newblock {\em Physica A: Statistical Mechanics and its Applications},
  637:129552, 2024.

\bibitem{A36}
Akanda Wahid-Ul-Ashraf, Marcin Budka, and Katarzyna Musial.
\newblock How to predict social relationships: Physics-inspired approach to
  link prediction.
\newblock {\em Physica A: Statistical Mechanics and its Applications},
  523:1110--1129, 2019.

\bibitem{A37}
Y.~V. Nandini, T.~Jaya~Lakshmi, Murali~Krishna Enduri, and Hemlata Sharma.
\newblock Link prediction in complex networks using average centrality-based
  similarity score.
\newblock {\em Entropy}, 26(6):433, 2024.

\bibitem{A40}
Yong-Jian He, Yijun Ran, Zengru Di, Tao Zhou, and Xiao-Ke Xu.
\newblock Uncovering multi-order popularity and similarity mechanisms in link
  prediction by graphlet predictors.
\newblock arXiv:2408.09406 [cs.SI], 2024.

\bibitem{A39}
Abdelhamid Saifi, Farid Nouioua, Mourad Charikhi, and Mouhoub Belazzoug.
\newblock Predicting missing links in complex networks using information about
  common neighbors and a degree of popularity.
\newblock {\em Evolving Systems}, 16(2):62, 2025.

\bibitem{A38}
Guangfu Chen, Bin Xie, and Yili Fang.
\newblock Link prediction in directed networks using hits centrality and biased
  random walks.
\newblock {\em Chaos, Solitons \& Fractals}, 200:116940, 2025.

\bibitem{A41}
Y.~V. Nandini, T.~Jaya~Lakshmi, Murali~Krishna Enduri, and Mohd Zairul~Mazwan
  Jilani.
\newblock Link prediction in complex hyper-networks leveraging hypercentrality.
\newblock {\em IEEE Access}, 13:12239--12254, 2025.

\bibitem{A6}
Marcus Engsig, Alejandro Tejedor, Yamir Moreno, Efi Foufoula-Georgiou, and
  Chaouki Kasmi.
\newblock Domirank centrality reveals structural fragility of complex networks
  via node dominance.
\newblock {\em Nature Communications}, 15(1):56, 2024.

\bibitem{A32}
Maksim Kitsak, Lazaros~K. Gallos, Shlomo Havlin, Fredrik Liljeros, Lev Muchnik,
  H.~Eugene Stanley, and Hern{\'a}n~A. Makse.
\newblock Identification of influential spreaders in complex networks.
\newblock {\em Nature Physics}, 6(11):888--893, 2010.

\bibitem{A33}
Jon~M. Kleinberg.
\newblock Authoritative sources in a hyperlinked environment.
\newblock {\em Journal of the ACM}, 46(5):604--632, 1999.

\bibitem{A34}
Linton~C. Freeman.
\newblock A set of measures of centrality based on betweenness.
\newblock {\em Sociometry}, 40(1):35--41, 1977.

\bibitem{A35}
Phillip Bonacich.
\newblock Power and centrality: A family of measures.
\newblock {\em American Journal of Sociology}, 92(5):1170--1182, 1987.

\bibitem{A29}
Xiyuan Wang, Haotong Yang, and Muhan Zhang.
\newblock Neural common neighbor with completion for link prediction.
\newblock In {\em Proceedings of the 12th International Conference on Learning
  Representations (ICLR)}, 2024.

\bibitem{A31}
Baolong Bi, Shenghua Liu, Yiwei Wang, Lingrui Mei, and Xueqi Cheng.
\newblock Lpnl: Scalable link prediction with large language models.
\newblock In {\em Findings of the Association for Computational Linguistics:
  ACL 2024}, pages 3615--3625, 2024.

\bibitem{A1}
Lawrence Page, Sergey Brin, Rajeev Motwani, and Terry Winograd.
\newblock The pagerank citation ranking: Bringing order to the web.
\newblock {\em Stanford Digital Libraries Working Paper}, 1998.

\bibitem{A43}
Elahe Nasiri, Kamal Berahmand, Zeynab Samei, and Yuefeng Li.
\newblock Impact of centrality measures on the common neighbors in link
  prediction for multiplex networks.
\newblock {\em Big Data}, 10(2):138--150, 2022.

\bibitem{A42}
Said Kerrache, Ruwayda Alharbi, and Hafida Benhidour.
\newblock A scalable similarity-popularity link prediction method.
\newblock {\em Scientific Reports}, 10(1):6394, 2020.

\bibitem{A44}
Jure Leskovec and Andrej Krevl.
\newblock Snap datasets: Stanford large network dataset collection.
\newblock http://snap.stanford.edu/data, 2014.

\bibitem{A4}
Ron Kohavi.
\newblock A study of cross-validation and bootstrap for accuracy estimation and
  model selection.
\newblock In {\em Proceedings of the 14th International Joint Conference on
  Artificial Intelligence (IJCAI)}, volume~2, pages 1137--1143, 1995.

\bibitem{A7}
James~A. Hanley and Barbara~J. McNeil.
\newblock The meaning and use of the area under a receiver operating
  characteristic (roc) curve.
\newblock {\em Radiology}, 143(1):29--36, 1982.

\bibitem{A45}
Xinshan Jiao, Yuxin Luo, Yilin Bi, and Tao Zhou.
\newblock Impacts of data splitting strategies on parameterized link prediction
  algorithms.
\newblock {\em Physica A: Statistical Mechanics and its Applications},
  692:131545, 2026.

\bibitem{A46}
R.~Aiyappa, X.~Wang, M.~Kim, O.~C.~Seckin, Y.-Y.~Ahn, and S.~Kojaku.
\newblock Implicit degree bias in the link prediction task.
\newblock In {\em Proceedings of the 42nd International Conference on Machine
  Learning (ICML)}, volume 267 of {\em Proceedings of Machine Learning Research},
  pages 874--908, 2025. \emph{arXiv:2405.14985}.

\bibitem{A47}
Yingying Zhang and Chengye Zhao.
\newblock Theoretical analysis of DomiRank centrality: automorphism, entropy, and
  graph transformations.
\newblock \emph{arXiv:2610.00107}, 2026.

\bibitem{A48}
Yingying Zhang and Chengye Zhao.
\newblock Block structure is invisible to the mean-field dismantling curve:
  degree-baseline fragility as a finite-size bridge-budget effect.
\newblock Companion manuscript, submitted (2026); preprint available from the
  corresponding author upon request.

\bibitem{A49}
Min Li, Shuming Zhou, and Gaolin Chen.
\newblock Link prediction of complex networks based on local path and closeness
  centrality.
\newblock In {\em Computational Data and Social Networks (CSoNet 2022)},
  volume 13831 of {\em Lecture Notes in Computer Science}, pages 51--63.
  Springer, 2023. \emph{DOI:10.1007/978-3-031-26303-3\_5}.

\bibitem{A50}
Haifeng Liu, Zheng Hu, Hamed Haddadi, and Hui Tian.
\newblock Hidden link prediction based on node centrality and weak ties.
\newblock {\em Europhysics Letters}, 101:18004, 2013.
  \emph{DOI:10.1209/0295-5075/101/18004}.

\bibitem{A51}
Serhan Y{\i}lmaz, Kaan Yorgancioglu, and Mehmet Koyut{\"u}rk.
\newblock Bias-aware training and evaluation of link prediction algorithms in
  network biology.
\newblock {\em Proceedings of the National Academy of Sciences}, 122(24):e2416646122,
  2025. \emph{DOI:10.1073/pnas.2416646122}.

\bibitem{A52}
Yu Wang and Tyler Derr.
\newblock Degree-related bias in link prediction.
\newblock In {\em 2022 IEEE International Conference on Data Mining Workshops
  (ICDMW)}, pages 757--758. IEEE, 2022.
  \emph{DOI:10.1109/ICDMW58026.2022.00103}.

\end{thebibliography}
\end{document}